\documentclass[preprints,review,accept,pdftex,moreauthors]{Definitions/mdpi} 
\firstpage{1} 
\pubvolume{2}
\issuenum{}
\articlenumber{}
\pubyear{2022}
\copyrightyear{2022}
\hreflink{https://doi.org/10.3390/cryst12121689} 
\doinum{10.3390/cryst12121689}
\pdfoutput=1

\Title{Ingredients for Generalized Models of $\kappa$-Phase Organic Charge-Transfer Salts: A Review}

\TitleCitation{Ingredients for Generalized Models of $\kappa$-Phase Organic Charge-Transfer Salts: A Review}

\Author{Kira Riedl, Elena Gati $^{2}$ and Roser Valent\'i $^{1}$}

\AuthorNames{Kira Riedl, Elena Gati and Roser Valent\'i}

\AuthorCitation{Riedl, K.; Gati, E.; Valent\'i, R.}

\address{%
$^{1}$ \quad Institut f\"ur Theoretische Physik, Goethe-Universit\"at Frankfurt, 60438 Frankfurt am Main, Germany; valenti@itp.uni-frankfurt.de \\
$^{2}$ \quad Max Planck Institute for Chemical Physics of Solids, 01187 Dresden, Germany; elena.gati@cpfs.mpg.de}

\corres{Correspondence: riedl@itp.uni-frankfurt.de}

\abstract{
The families of organic charge-transfer salts $\kappa$-(BEDT-TTF)$_2X$ and $\kappa$-(BETS)$_2X$, where BEDT-TTF and BETS stand for the organic donor molecules C$_{10}$H$_8$S$_8$ and C$_{10}$H$_8$S$_4$Se$_4$, respectively, and $X$ for an inorganic electron acceptor, have been proven to serve as a powerful playground for the investigation of the physics of frustrated Mott insulators. These materials have been ascribed a model character, since the dimerization of the organic molecules allows to map these materials onto a single band Hubbard model, in which the dimers reside on an anisotropic triangular lattice. By changing the inorganic unit $X$ or applying physical pressure, the correlation strength and anisotropy of the triangular lattice can be varied.  This has led to the discovery of a variety of exotic phenomena, including quantum-spin liquid states, a plethora of long-range magnetic orders in proximity to a Mott metal-insulator transition, and unconventional superconductivity.
While many of these phenomena can be described within this effective one-band Hubbard model on a triangular lattice, it has become evident in recent years that this simplified
description is insufficient to capture all observed magnetic and electronic properties. The ingredients for generalized models that are relevant include, but are not limited to, spin-orbit coupling, intra-dimer charge and spin degrees of freedom,  electron-lattice coupling, as well as disorder effects. Here, we review selected theoretical and experimental discoveries that clearly demonstrate the relevance thereof. At the same time, we outline that these aspects are not only relevant to this class of organic charge-transfer salts, but are also receiving increasing attention in other classes of inorganic strongly correlated electron systems.
This reinforces the model character that the $\kappa$-phase organic charge-transfer salts have for understanding and discovering novel phenomena in strongly correlated electron systems from a theoretical and experimental point of view.}


\begin{document}

\section{Introduction}
\label{sec:introduction}

\subsection{Crystal and Electronic Structure}

\begin{figure}[H]
\centering
\includegraphics[width=0.99\textwidth]{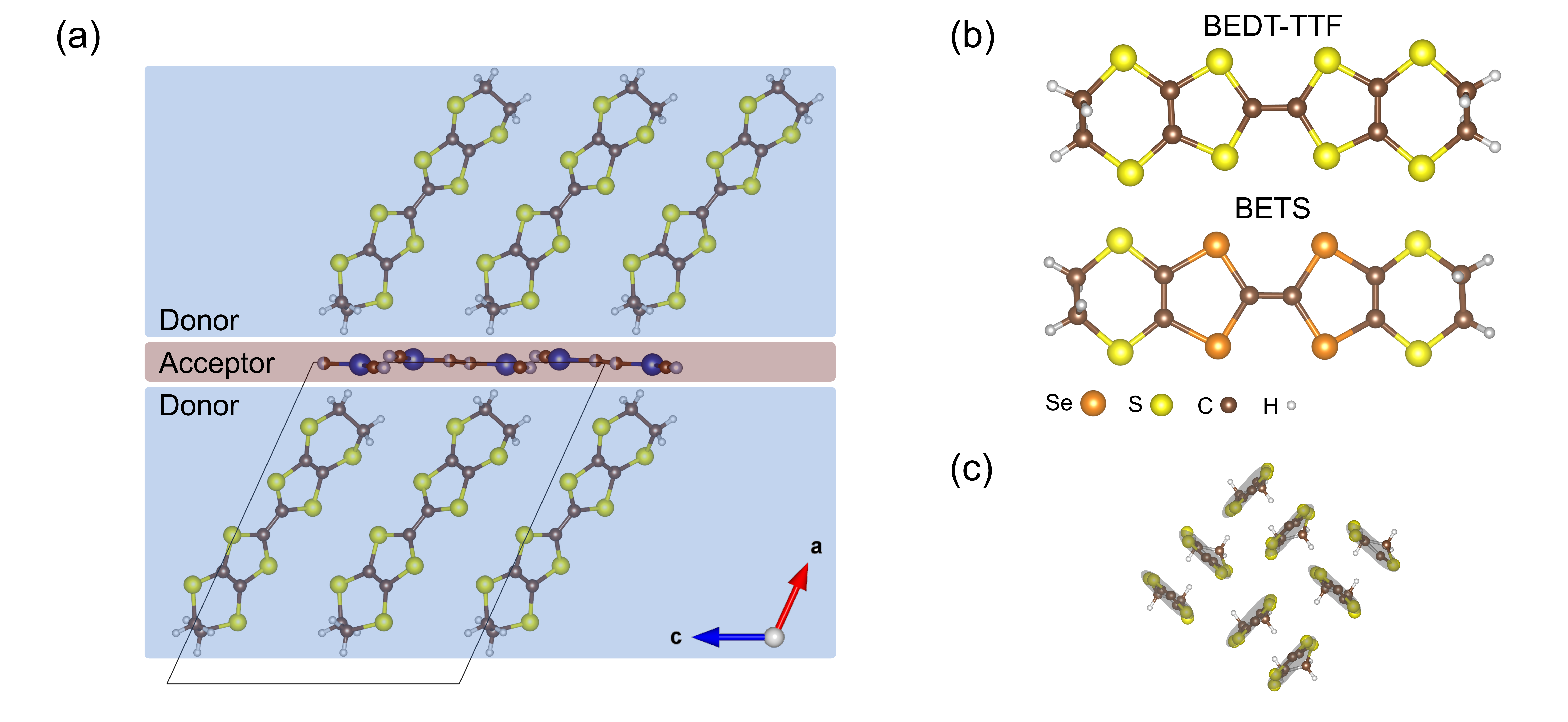}
\caption{(\textbf{a}) Layered structure of organic electron donor and inorganic electron acceptor layer, shown for the example $\kappa$-(BEDT-TTF)$_2$Cu$_2$(CN)$_3$; (\textbf{b}) structure of BEDT-TTF and BETS molecules; and (\textbf{c})~\mbox{$\kappa$ packing} motif of the organic layer of $\kappa$ phase charge-transfer salts from the top view. 
}
\label{fig:crystal-structure}
\end{figure}

Organic charge-transfer salts have received, in the last decades, a lot of attention for their variability as materials and their model role in understanding most of the challenging phenomena in correlated systems \cite{Kanoda97,Toyota07}, such as the Mott-metal insulator transition \cite{Limelette03b,Fournier03,Pustogow21}, Mott criticality \cite{Kagawa09,Furukawa15b,Isono16,Gati17}, quantum-spin liquid phases \cite{Shimizu03,Powell11,Riedl19,Pustogow22},
unconventional superconductivity \cite{wosnitza19,Guterding16b} or multiferroicity \cite{Lunkenheimer12}, among others. One of the most studied families is the $\kappa$ phase. The $\kappa$ phase in organic charge-transfer salts consists of alternation in charge donating organic BEDT-TTF {(BEDT-TTF = bisethyelenedithio-tetrathiafulvalene)} or BETS {(BETS = bisethylenedithio-tetraselenafulvalene)} layers and 
acceptor inorganic anion layers, as depicted in Figure~\ref{fig:crystal-structure} for the example of $\kappa$-(BEDT-TTF)$_2$Cu$_2$(CN)$_3$ crystallizing in the space group $P2_1/c$. 

\begin{figure}[H]
\centering
\includegraphics[width=0.98\textwidth]{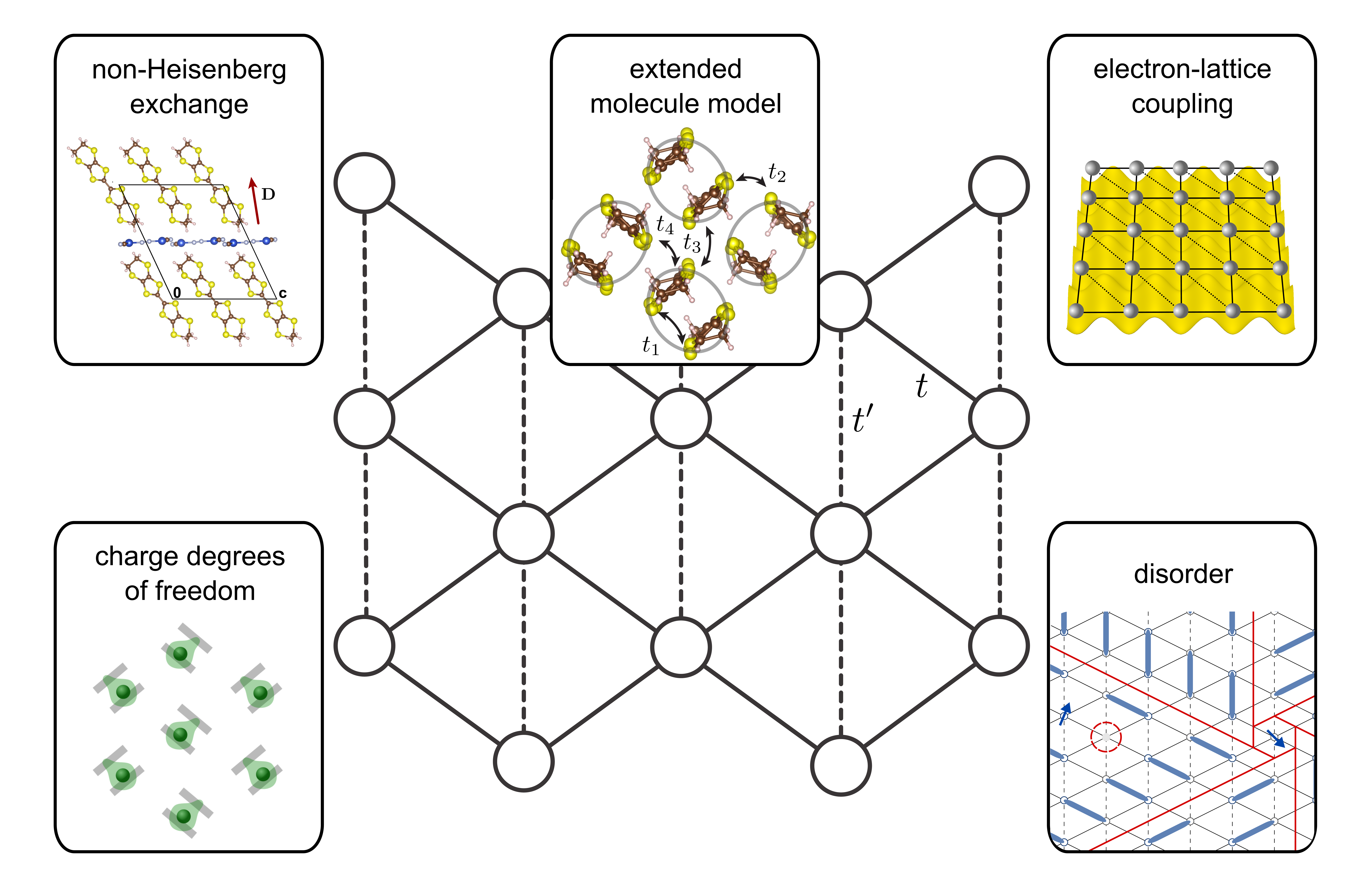}
\caption{Center: Illustration of the mapping of the organic layer 
of $\kappa$ phase charge-transfer salts to an anisotropic triangular lattice. 
The various ingredients for generalized models beyond the strongly dimerized one-band Hubbard picture discussed in this work are illustrated in the boxes.
}
\label{fig:overview}
\end{figure}

The choice of different inorganic layers $X$ in $\kappa$-(BEDT-TTF)$_2X$ and $\kappa$-(BETS)$_2X$ influences the orbital overlap between the organic molecules and allows for distinct magnetic and electronic properties in these systems.  In the charge-transfer process, each molecule donates  a charge of half an electron.
The $\kappa$-type arrangement of organic molecules exhibits
a strong dimerization of the molecules forming 
a triangular lattice of dimers with one hole per dimer, as illustrated in 
Figures~\ref{fig:crystal-structure}c and \ref{fig:overview}.

Alternatively, one can consider the molecules within a dimer
 as the building blocks (see inset in Figure~\ref{fig:overview}) 
 leading to an extended molecule model with half a hole per molecule.
  By symmetry, there are four distinct hopping parameters between the
 highest occupied molecular orbitals (HOMO) $\vert g_i \rangle$ of the molecules.
 The hoppings are conventionally abbreviated as $t_{1 \ldots 4}$, as shown in Figure~\ref{fig:overview}. 
 Here, $t_1$ is the intra-dimer hopping
 and one can make use of geometrical expressions $t=(t_2+t_4)/2$ and $t^\prime = t_3/2$ \cite{Komatsu96,Kandpal09}
 to relate the $t_{1\ldots 4}$ hoppings of
 the extended molecule model to the $t$, $t^\prime$ hoppings of the dimer model shown in Figure~\ref{fig:overview}.
 The corresponding electronic structure of the dimer model consists, then, of an anti-bonding dimer
 orbital [$\vert a \rangle = \frac{1}{\sqrt{2}}(\vert g_1 \rangle + \vert g_2 \rangle )$], occupied by one electron (or one hole), and of a bonding dimer orbital [$\vert b \rangle = \frac{1}{\sqrt{2}}(\vert g_1 \rangle - \vert g_2 \rangle )$] occupied by two electrons.

\subsection{Phase Diagram}
\label{sec:phase-diagram}

Under the assumption of very strong dimerization, it is well-established, as mentioned above, that the $\kappa$-phase organic charge-transfer salts are model systems to realize the half-filled one-band Hubbard model \cite{Kanoda97,Toyota07,Kandpal09,Powell11,Hotta12,Pustogow18}:
\begin{linenomath}
\begin{equation}
\mathcal{H}\,=\,-\sum_{<i,j>,\sigma} (t_{ij} c_{i,\sigma}^{\dagger} c_{j,\sigma} + h.c.) + \sum_{i} U \, n_{i,\uparrow} n_{i,\downarrow}
\label{eq:Hubbard}
\end{equation}
\end{linenomath}
with hopping $t_{ij} = t, t^\prime$ (see Figure~\ref{fig:overview}).
One key parameter of this model is the ratio of the strength of Coulomb repulsion $U$ to the kinetic energy  $W\,\sim\,t,t^\prime$.  The $\kappa$-phase charge-transfer salts all lie in a range close to the Mott transition, i.e., $U/W\,\sim\,1$. 

\begin{figure}[t]
\centering
\includegraphics[width=0.85\textwidth]{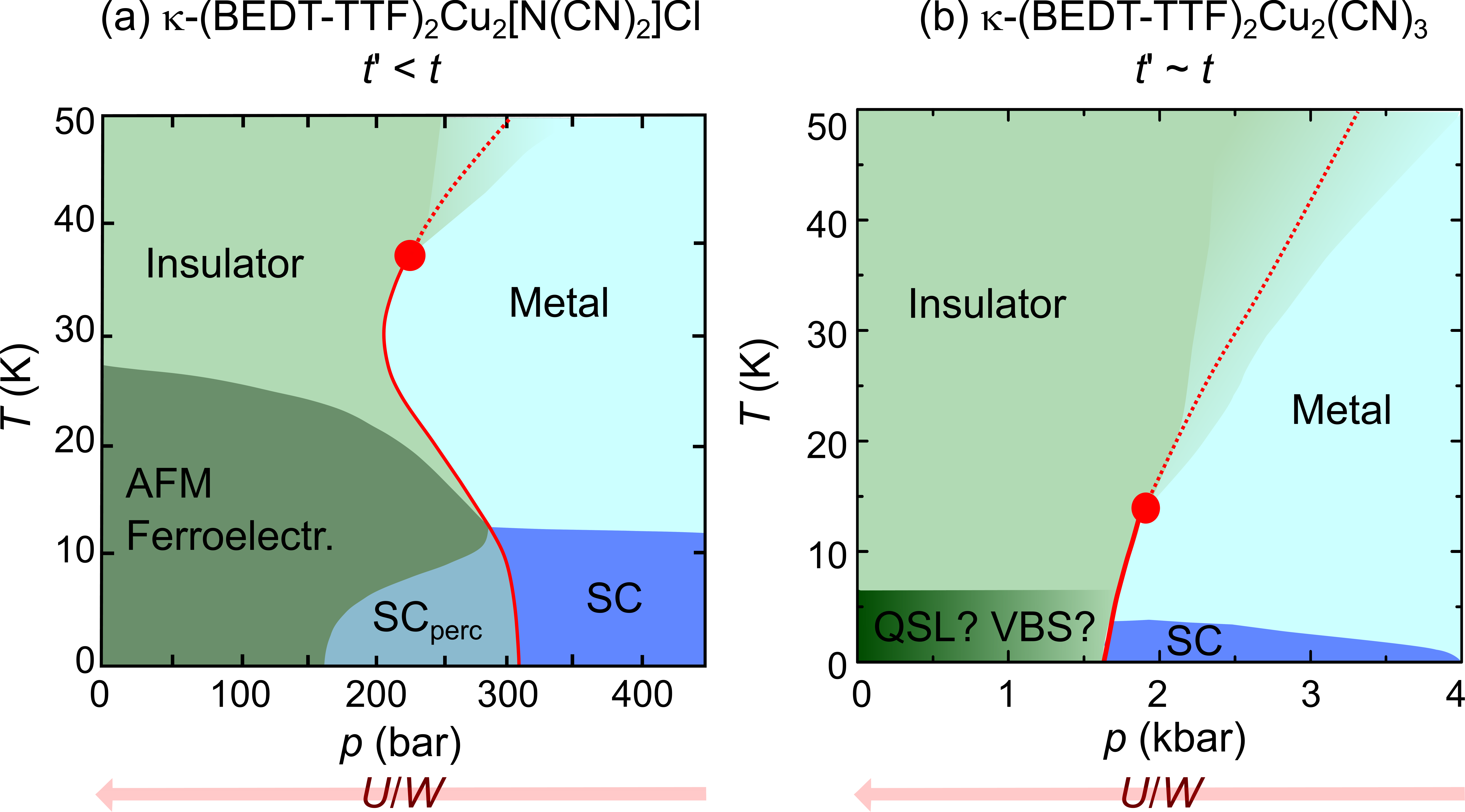} 
\caption{Experimental temperature--pressure phase diagrams of $\kappa$-phase charge-transfer salts: (\textbf{a})~$\kappa$-(BEDT-TTF)$_2$Cu[N(CN)$_2$]Cl (‘$\kappa$-Cl') \cite{Kanoda97,Lefebvre00,Limelette03b,Fournier03,Kagawa04,Lunkenheimer12,Gati17} and (\textbf{b}) $\kappa$-(BEDT-TTF)$_2$Cu$_2$(CN)$_3$ \linebreak (‘$\kappa$-CuCN')~\cite{Shimizu03,Kurosaki05,Miksch21,Furukawa15b}. At ambient pressure and low-enough temperatures, both compounds exhibit a Mott insulating ground state (i.e., a moderate to large correlation strength, $U/W$). Upon applying pressure, $U/W$ is reduced and eventually a first-order Mott metal-insulator transition is induced (red line) which ends in a second-order critical endpoint (red circle). On both sides of the Mott transition, intriguing electronic orders emerge at lowest temperatures. On the Mott insulating side, $\kappa$-Cl orders antiferromagnetically \cite{Miyagawa95}, which is well-understood given the anisotropy of its triangular lattice ($t'< t$). The antiferromagnetic (AFM) order is believed to be accompanied by the emergence of long-range ferroelectric order, making this compound multiferroic \cite{Lunkenheimer12}. In contrast, $\kappa$-CuCN is characterized by an almost isotropic triangular lattice with $t'\,\sim\,t$. For a long time, it has, therefore, been considered as a candidate for quantum spin-liquid (QSL) behavior  \cite{Shimizu03}. However, there exist recent results which argue in favor of the formation of a valence bond solid (VBS) \cite{Riedl19,Pustogow20,Miksch21,Matsuura22,Pustogow22,Pustogow22b}. On the metallic side of the Mott transition, both compounds exhibit superconductivity (SC). A region of percolative superconductivity SC$_{\rm perc}$ can also be found in the Mott insulating state close to the metal-insulator boundary.}
\label{fig:pressurephasediagram}
\end{figure}

\textls[-25]{The essential features of the Hubbard model can be identified
in experimental {\it temperature -- pressure} ($T-p$) phase diagrams of
$\kappa$-phase organic charge-transfer salts. The application of hydrostatic pressure causes an increase in the orbital overlap and, therefore, an increase in $W$ (or decrease in $U/W$). As a result, a Mott insulator is expected to undergo an insulator-to-metal transition. This is indeed observed for various Mott-insulating $\kappa$-phase 
charge-transfer salts, including \linebreak \mbox{$\kappa$-(BEDT-TTF)$_2$Cu[N(CN)$_2$]Cl} \mbox{\cite{Kanoda97,Lefebvre00,Limelette03b,Fournier03,Kagawa04,Gati17}} (‘$\kappa$-Cl', Figure \ref{fig:pressurephasediagram}a) and $\kappa$-(BEDT-TTF)$_2$Cu$_2$(CN)$_3$ \cite{Kurosaki05,Furukawa15b,Pustogow21}} \textls[-50]{(‘$\kappa$-CuCN', Figure\,\ref{fig:pressurephasediagram}b) and \mbox{$\kappa$-(BETS)$_2$Mn[N(CN)$_2$]$_3$}} (\mbox{‘$\kappa$-Mn}') \cite{Zverev10,Zverev19,Vyaselev11}. In all cases, very moderate pressures in the order of kbar, or even less, induce a Mott metal-insulator transition (MIT). This finding demonstrates, on the one hand, the proximity of various $\kappa$-phase charge-transfer salts to the Mott MIT and, on the other hand, it also shows the  tunability of these materials in laboratory settings through physical and chemical pressures \cite{Kanoda97,Toyota07}.

At low temperatures, the pressure-induced Mott MIT in $\kappa$-Cl and $\kappa$-CuCN is found to be a first-order transition. Upon increasing temperature, the first-order line ends in a second-order critical endpoint at $(p_{cr},T_{cr})$. Above the endpoint, only a crossover but no phase transition exists, similar to the liquid--gas transition. In the purely electronic model of the Mott transition, the Mott critical endpoint is, thus, expected to fall into the Ising universality class \cite{Castellani79,Kotliar00}. This notion of Mott criticality is further corroborated by DMFT (dynamical mean field theory) calculations \cite{Georges96}, which are also able to predict the order of the Mott MIT.  In Section\,\ref{sec:lattice}, we will discuss the limitations of this purely electronic model when the coupling to lattice degrees of freedom becomes relevant.

Whereas the behavior very close to the first-order critical endpoint at $T_{\rm cr}$ is one of a classical phase transition, an intriguing observation in $\kappa$-phase charge-transfer salts is a quantum-critical scaling of measured
transport in the crossover region $T$\,$\gg$\,$T_{\rm cr}$. Such a quantum-critical scaling was predicted based on DMFT calculations of the Hubbard model~\cite{Terletska11,Vucicevic13}. 
 While one would typically predict quantum-critical scaling to occur at lowest temperatures $T\,\rightarrow\,$0, it is important to note that the dominant energy scales \cite{Pustogow18} of the Mott transition ($U$ and $W$) are of the order of 1000\,K\,$\gg\,T_{\rm cr}$ . Thus, for intermediate temperatures, the system effectively behaves as if the Mott critical endpoint were located at zero. In fact, frustration suppresses magnetic ordering tendencies, so that frustrated organic charge-transfer salts are suggested to show the properties of genuine Mott systems to lower temperatures \cite{ferber2014,Pustogow18}. Such an importance of magnetic frustration for investigating paramagnetic Mott metal-insulating transitions has been recently discussed as well, in the context of inorganic V$_2$O$_3$ \cite{leiner2019}. 

The  spin degrees of freedom in the Mott insulating state of  $\kappa$-phase charge-transfer salts become important at very low temperatures $T\,\rightarrow\,0$. $\kappa$-Cl undergoes a transition into an antiferromagnetic (AFM) ordered ground state at $T_N\,\sim\,$27\,K \cite{Miyagawa95}. In the dimer model, AFM order might be expected when the triangular lattice is very anisotropic and close to the square lattice, which is indeed the case for $\kappa$-Cl ($t^\prime/t \approx 0.45$~\cite{Kandpal09,Koretsune14}). In contrast, when the triangular lattice is more isotropic, as is the case for $\kappa$-CuCN ($t^\prime/t \approx$ 0.8--0.9~\cite{Kandpal09,Nakamura09,Nakamura12,Jeschke12,Koretsune14}), various theoretical studies
of the Hubbard model on the triangular lattice at half-filling 
suggest the emergence of a quantum spin liquid, as, for instance, in Refs. \cite{tocchio2013,oleg2015,szasz2020}. In fact, $\kappa$-CuCN has been considered for a long time as a prime candidate for the realization of 
a quantum spin liquid \cite{Shimizu03,Yamashita08} that potentially hosts gapless excitations \cite{Lee05}. 
However, thermal expansion measurements \cite{Manna12,Hartmann18} indicate the presence of a phase transition around 6\,K, often referred to as the ``6\,K anomaly'', which might prevent the formation of a possible spin-liquid phase. In recent years, the interpretation of magnetic torque~\cite{Isono16} and NMR~\cite{Shimizu06} data 
suggested that a valence-bond solid rather than a quantum spin liquid might be formed in $\kappa$-CuCN~\cite{Riedl19} below $T^\ast\,\sim\,6\,$K. This picture was further supported by recent ESR
studies confirming the presence of a spin gap opening~\cite{Miksch21} around the 6\,K anomaly. Still, open questions regarding the magnetic ground state of $\kappa$-CuCN remain; see Ref. \cite{Pustogow22} for a recent review of the large set of experimentally available data on this system.

$\kappa$-Mn also shows magnetic order for $T\,<\,30$\,K in the Mott insulating state at the lowest temperature \cite{Zverev10,Vyaselev11,Riedl21}. However, the magnetic order that is realized in $\kappa$-Mn is not of the N\'eel type, as observed in $\kappa$-Cl. In fact, it was shown that $\kappa$-Mn realizes a spin-vortex crystal order that can only be understood when taking ring-exchange interactions into account~\cite{Riedl21}, as we will explain in detail in Section\,\ref{sec:non-Heisenberg}. In addition, in contrast to $\kappa$-Cl, where the magnetic ordering occurs within the Mott insulating state, the onset of magnetic order in $\kappa$-Mn coincides with a first-order metal-to-Mott insulator transition at \mbox{$T_{\rm MI}\,=\,T_{\rm N}$~\cite{Zverev10}}. The ordered, low-entropy state in $\kappa$-Mn dictates a negative slope of d$T_{\rm MI}$/d$p$ in the temperature--pressure phase diagram; see also Refs.\,\cite{Pustogow22,Pustogow22b} for discussions of d$T_{\rm MI}$/d$p$ in $\kappa$-Cl and~\mbox{$\kappa$-CuCN}.

On the metallic side of the Mott MIT, superconductivity is often found at low temperatures \cite{Lang96,Wosnitza07}, as seen in the temperature--pressure phase diagrams of $\kappa$-Cl, $\kappa$-CuCN and $\kappa$-Mn, with moderate transition temperatures $T_{\rm c}\sim\,10$\,K (see Figure~\ref{fig:pressurephasediagram}). Note that also chemical modifications, e.g., replacing Cl completely by Br in $\kappa$-Cl, can be used to obtain superconducting metals at ambient pressure. The proximity of the superconducting phase to a (magnetic) Mott insulating phase suggests a close connection between the $\kappa$-phase organic
charge-transfer salts and the  high-temperature cuprate superconductors \cite{McKenzie97}, where the origin
of superconductivity is attributed to the presence of spin fluctuations.  Actually,
a few theoretical treatments of the dimer model (see, e.g., Refs. \cite{Schmalian98,Kino98,kyung2006,Hebert15,Watanabe06,Guterding16,Guterding16b,Zantout18}) follow the
scenario of superconductivity mediated by spin fluctuations.

\subsection{Outline of This Review}

Whereas the one-band Hubbard model on the triangular
lattice clearly captures essential features of the phase diagram of the $\kappa$-phase organic charge-transfer salts, it has become evident in recent years that generalized models for $\kappa$-phase organic charge-transfer salts must contain additional contributions to accurately account for all salient features of their phase diagrams. The present review summarizes theoretical and experimental works, based on which the importance of specific, additional interactions were suggested. The paper is structured as follows (see also Figure\,\ref{fig:overview} for a sketch of the synopsis). In Section\,\ref{sec:non-Heisenberg}, we discuss the role of spin-orbit coupling and higher order four-spin ring exchange couplings for the magnetic properties of the Mott insulating $\kappa$-phase charge-transfer salts. We then proceed, in Section\,\ref{sec:molecule-based}, with a summary of new effects in the extended molecule model such as ferroelectricity in the Mott insulating state or mixed superconducting order parameters. In the following Section\,\ref{sec:lattice}, we present evidence for the importance of electron-lattice coupling in these correlated electron systems. The last aspect that we will cover in Section\,\ref{sec:disorder} is the role of disorder for the properties of the $\kappa$-phase charge-transfer salts close to the Mott transition.
In Section\,\ref{sec:summary}, we present a conclusion and outlook of the review and discuss specific important open questions and their relevance for the broader field of correlated electron systems.

\section{Magnetic Exchange beyond Heisenberg}
\label{sec:non-Heisenberg}

\begin{figure}[H]
\includegraphics[width=\columnwidth]{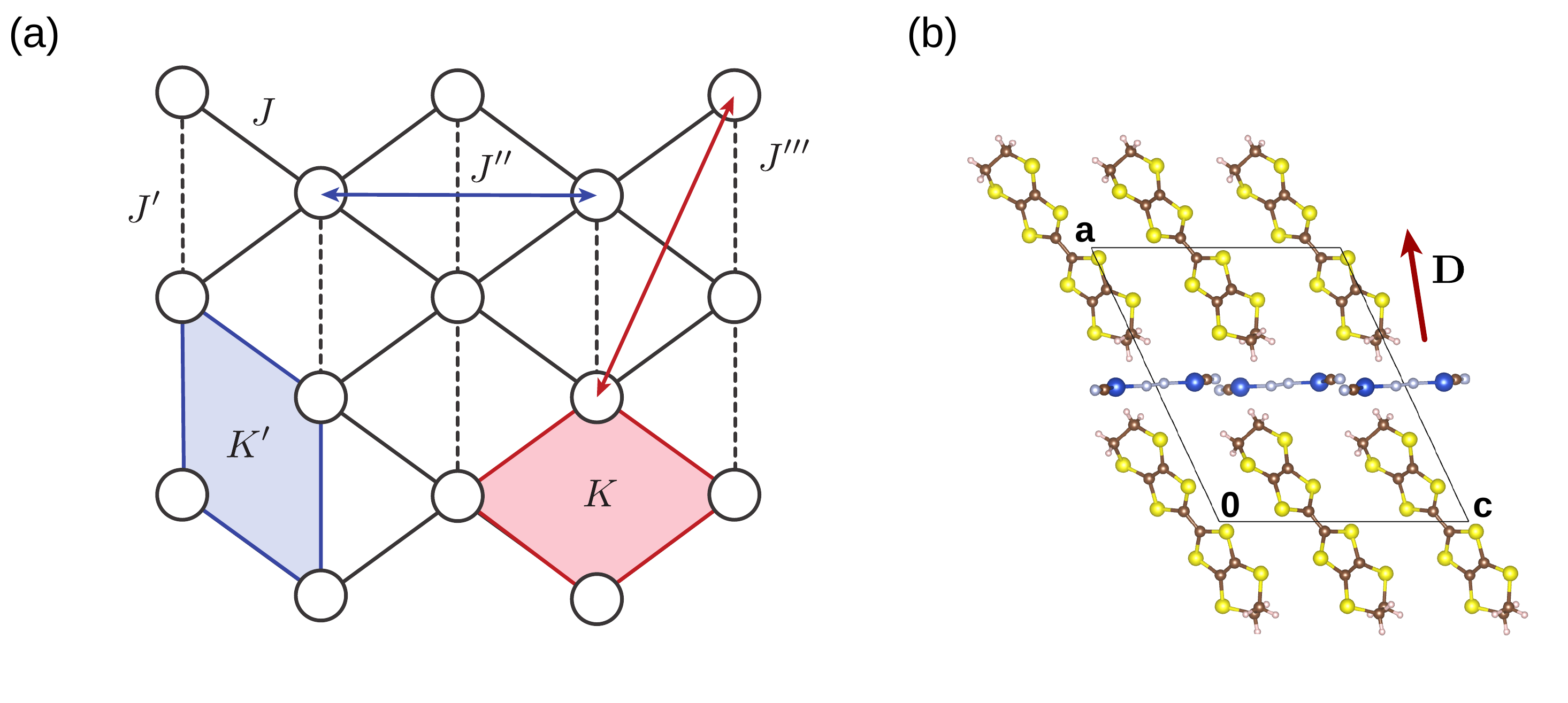} 
\vspace{-6pt} 
\caption{Magnetic exchange in $\kappa$-phase charge-transfer salts. (\textbf{a}) Heisenberg exchange terms $J$, $J^\prime$, $J^{\prime \prime}$, $J^{\prime \prime \prime}$ and four-spin ring-exchange $K$ and $K^\prime$ on the two distinct four-site plaquettes on the anisotropic triangular lattice.
(\textbf{b}) The SOC-induced DM vector $\mathbf{D}$ is oriented approximately along the long side of the organic molecule, here illustrated for the example $\kappa$-(BEDT-TTF)$_2$Cu$_2$(CN)$_3$.
Figures reprinted from Refs.~\cite{Riedl21,Winter17importance}.}
\label{fig:non-Heisenberg}
\end{figure}

Deep in the Mott insulating phase,
the charge degrees of freedom do not influence the low energy properties of the system
and a description in terms of a purely spin-1/2 magnetic model on the triangular lattice 
is a suitable starting point, where one dimer represents one magnetic site 
(see Figure~\ref{fig:non-Heisenberg}).
The dominant magnetic exchange terms are of the Heisenberg type:
\begin{linenomath}
\begin{equation}
\mathcal{H} = \sum_{\langle ij \rangle} \,J_{ij} \, \mathbf{S}_i \cdot \mathbf{S}_j,
\end{equation}
\end{linenomath}
with exchange $J$ on bonds indicated by solid lines and $J^\prime$ on bonds indicated by dashed lines in Figure~\ref{fig:non-Heisenberg}a. This model can be obtained from a
perturbation expansion of the Hubbard model (Equation~\eqref{eq:Hubbard}) in powers of $t/U$~\cite{oles78,gros87}.
Depending on the ratio $J^\prime/J$, it is possible to scale in between the limit of a square lattice ($J^\prime/J=0$), an isotropic triangular lattice ($J^\prime/J=1$) and one-dimensional chains ($J^\prime/J=\infty$). In the $\kappa$-phase charge-transfer salts, this ratio is determined by the nature of the anion layer.
Further Heisenberg exchanges up to fourth neighbors are shown in  Figure~\ref{fig:non-Heisenberg}a.

Beyond the Heisenberg exchange there are two main contributions, which, albeit being subdominant, may change the nature of the magnetic ground state of
the $\kappa$-phase charge-transfer salts completely. One is the spin-orbit coupling, which allows for the presence of anisotropic bilinear magnetic exchange, such as the Dzyaloshinskii--Moriya vector $\bf{D}$ (Figure~\ref{fig:non-Heisenberg}b), changing the symmetry properties of the magnetic model severely. A second important contribution is the four-spin ring exchange $K$, $K^\prime$, which is expected to be non-negligible for materials close to the Mott MIT.
Note that in the presence of a magnetic field, additional terms with an odd number of spins, such as the scalar spin chirality, are also present. We do not discuss these terms here.
\vspace{-6pt} 

\subsection{Magnetic Bilinear Anisotropic Interactions}

Magnetic anisotropic interactions arise due to the presence of spin-orbit coupling. 
Originally, this aspect was often ignored in the context of the organic charge-transfer salts due to the light nature of the Se, S, C and H atoms in the BEDT-TTF and BETS molecules.
However, it was pointed out early on \cite{Balents10} that at least at very low temperatures the influence of anisotropic interactions may be significant to capture new magnetic phenomena. The full bilinear magnetic model for $S=1/2$ can be expressed as follows:
\begin{linenomath}
\begin{equation} \label{eq:bilinear_Hamiltonian}
\mathcal{H}_{(2)} = \sum_{\langle ij \rangle} J_{ij} \, \mathbf{S}_i \cdot \mathbf{S}_j 
     + \mathbf{D}_{ij} \cdot (\mathbf{S}_i \times \mathbf{S}_j)
     + \mathbf{S}_i \cdot \mathbf{\Gamma}_{ij} \cdot \mathbf{S}_j,
\end{equation}
\end{linenomath}
with the antisymmetric Dzyaloshinskii--Moriya (DM) vector $\mathbf{D}_{ij}$ and the symmetric pseudo-dipolar tensor $\mathbf{\Gamma}_{ij}$. Note that on the bonds indicated by the dashed lines in Figure~\ref{fig:non-Heisenberg}a, the DM vector vanishes due to an inversion center at the bond center. 
To obtain an intuition about the strengths of the anisotropic terms, we consider perturbation-theory expressions in the strongly localized limit with weak SOC, i.e., $U\gg t,|\vec{\lambda}|$. Here, $\vec{\lambda}$ is a spin-dependent hopping term in the electronic Hubbard picture, arising due to the presence of SOC, $\mathcal{H}_{\rm hop}=\sum_{ij} \mathbf{c}_{i }^\dagger ( t_{i j } I_{2 \times 2} + \frac{i}{2} \vec{\lambda}_{i j } \cdot \vec{\sigma} ) \mathbf{c}_{j }$~\cite{bernevig2013}, with the single particle operator $\mathbf{c}^\dagger=(c_{i \uparrow}^\dagger \quad c_{i \downarrow}^\dagger )$ and the Pauli matrices $\vec{\sigma}$.
In second-order perturbation theory, the bilinear-exchange couplings scale with $J \propto t^2/U$, $ \mathbf{D} \propto (t\, \vec{\lambda})/ U$, and $ \mathbf{\Gamma}  \propto (\vec{\lambda} \otimes \vec{\lambda})/U$ \cite{Winter17importance}. In $\kappa$-phase charge-transfer salts, the hopping amplitude is an order of magnitude larger than SOC, so that the Heisenberg term is expected to be dominant, with important contributions of the DM vector at very low temperatures and negligible contributions of the pseudo-dipolar tensor. 

In Table~\ref{tab:exchange}, we list selected ab-initio results~\cite{Winter17importance,Riedl19,Riedl19thesis,Riedl21} for various $\kappa$-phase organic compounds. The nearest-neighbor Heisenberg-exchange couplings are on the order of a few hundred Kelvin, where the ratio $J^\prime/J$ indicates whether the material is closer to the square lattice limit ($J^\prime/J<1$) or whether it is approaching the limit of one-dimensional chains ($J^\prime/J>1$). The DM vector is, as expected, significantly smaller with $\vert \mathbf{D} \vert \sim 5\,$K for the BEDT-TTF and $\vert \mathbf{D} \vert \sim 25\,$K for the BETS compounds. The stronger anisotropic contribution in the BETS material can be directly related to the presence of the heavier Se atoms in the organic molecule (see Figure~\ref{fig:crystal-structure}b). The DM vector
is consistently oriented approximately
along the long side of the molecule, as shown in Figure~\ref{fig:non-Heisenberg}b for the example of $\kappa$-(BEDT-TTF)$_2$Cu$_2$(CN)$_3$. This implies that the main contribution is the component perpendicular to the triangular plane of molecular dimers. For all discussed Mott insulators in this review, this out-of-plane component has a staggered pattern, while the component along the dashed bonds in Figure~\ref{fig:non-Heisenberg}a has a stripy pattern by symmetry~\cite{Winter17importance,Riedl21}. 

\begin{table}[H]
\caption{Representative results for indicated compounds and reference for exchange parameters in K: Bilinear exchange $J$, $J^\prime$, $\mathbf{D}$ (defined in Eq.~\eqref{eq:bilinear_Hamiltonian} and Fig.~\ref{fig:non-Heisenberg}) and averaged four-spin ring exchange $K$, $K^\prime$ (defined in Eq.~\eqref{eq:ring-exchange_Hamiltonian} and Fig.~\ref{fig:non-Heisenberg}). The DM vectors for the $Pnma$ salts are given in the coordinate system $(a,b,c)$, while the $P21/c$ values are indicated with $()^\ast$ and given with
respect to $(a,b,c^\ast)$. The ratio $J^\prime/J$ indicates the deviation from the isotropic triangular limit $J^\prime/J=1$ and $K/J$ indicates the significance of the four-spin ring exchange. This table is not intended to be a complete representation of available results, but serves as orientation with selected values.
\label{tab:exchange}}
\newcolumntype{L}[1]{>{\hsize=#1\hsize\raggedright\arraybackslash}X}
\newcolumntype{R}[1]{>{\hsize=#1\hsize\raggedleft\arraybackslash}X}
\newcolumntype{C}[1]{>{\hsize=#1\hsize\centering\arraybackslash}X}
\begin{tabularx}{\textwidth}{L{4.35} |C{0.3} C{0.3} C{0.3} L{2.75}| C{0.3} C{0.3} C{0.3} | C{0.55}| C{0.55}}
\toprule
Compound 	& Ref.  &$J$ & $J^\prime$ & \centering{$\mathbf{D}$} & Ref. & $K$ & $K^\prime$ & $J^\prime/J$ & $K/J$\\
\midrule
$\kappa$-(BEDT-TTF)$_2$Cu[N(CN)$_2$]Cl 
& 
\cite{Winter17importance} 
& 482 & 165 
& $(-3.6,-3.6,-0.2)$ 
&
\cite{Riedl19thesis} 
& 62 & 21 
& 0.34 & 0.13 \\
$\kappa$-(BEDT-TTF)$_2$Ag$_2$(CN)$_3$ 
&\cite{Winter17importance}
& 250 & 158
& $(-2.9,-0.9,-2.9)^\ast$ 
&
\cite{Riedl19thesis}
& 20 & 13
& 0.63 & 0.08 \\
$\kappa$-(BEDT-TTF)$_2$Cu$_2$(CN)$_3$
&\cite{Winter17importance} 
& 228 & 268 
& $(+3.3,+0.9,+1.0)^\ast$  
&
\cite{Riedl19}
& 18 & 18
& 1.18 & 0.08 \\
$\kappa$-(BEDT-TTF)$_2$B(CN)$_4$ 
&
\cite{Winter17importance} 
& 131 & 366 
& $(+1.0,+4.2,-0.1)$ 
&
\cite{Riedl19thesis} 
& 5 & 15
& 
2.79 & 0.04\\
$\kappa$-(BETS)$_2$Mn[N(CN)$_2$]$_3$
& 
\cite{Riedl21}
& 260 & 531
& $(+22.6, -1.9,+ 8.8)^\ast$
&
\cite{Riedl21}
& 16 & 39
& 
2.04 & 0.06\\
\bottomrule
\end{tabularx}
\end{table}

In the well-studied material $\kappa$-(BEDT-TTF)$_2$Cu[N(CN)$_2$]Cl, the presence of the DM interaction can be directly related to an observed small ferromagnetic contribution to an antiferromagnetic transition at $T_{\rm N}=27\,$K by ESR~\cite{Miyagawa95}. This observation is consistent with a weak canting of the magnetic moments, as one would expect from the presence of a DM interaction in an AFM N\'eel ordered state.

Further important effects of bilinear anisotropic interactions can be
detected when the spin system
couples to an external magnetic field $\mathbf{H}$. To fully grasp the consequences of these anisotropic interactions, it is helpful to follow an approach first introduced in the 1990s~\cite{Shekhtman92}. Specific DM patterns (e.g., uniform and staggered DM patterns) allow to apply local rotations for each spin, so that the anisotropic interactions may be ``gauged'' away. Note that for this approach, it is assumed that $\mathbf{\Gamma} \propto \mathbf{D} \otimes \mathbf{D}$, which is generally not perfectly fulfilled in the presence of Hund's coupling. 
While this condition is not fully obeyed, the general behavior of the system in the presence of an external field may still be described sufficiently with this model.
A local rotation about the angle $\eta_i \Phi$, with $\eta_i=\pm1$ for different sublattices and $\Phi = 1/2 \arctan (J/|\mathbf{D}|)$, results, then, in a bilinear isotropic Heisenberg Hamiltonian in the presence of a uniform and of a staggered magnetic field:

\begin{linenomath}
\begin{equation}
    \mathcal{H}_{(2)} = \sum_{\langle ij \rangle} \tilde{J}_{ij} \,\tilde{\mathbf{S}}_i \cdot \tilde{\mathbf{S}}_j
- \mu_{\rm B} \sum_{i} (\mathbf{H}_{\rm eff,u} + \eta_i \mathbf{H}_{\rm eff,s}) \cdot \tilde{\mathbf{S}}_i.
\end{equation}
\end{linenomath}
Here, $\tilde{\mathbf{S}}$ are the rotated spin operators, $\mathbf{H}_{\rm eff,u} \approx \mathbb{G}_{\rm u}^{\rm T} \cdot \mathbf{H}$ is the effective uniform magnetic field, $\mathbf{H}_{\rm eff,s} \approx \mathbb{G}_{\rm s}^{\rm T} \cdot \mathbf{H} - \frac{1}{2J}((\mathbb{G}_{\rm u}^{\rm T} \cdot \mathbf{H}) \times \mathbf{D})$ the effective staggered magnetic field, and $\mathbb{G}_{\rm u/s}$ the uniform/staggered contribution to the gyromagnetic tensor. 
In other words, applying an external uniform field $\mathbf{H}$ to an anisotropic magnetic material with a staggered DM vector is equivalent to applying an external uniform ($\mathbf{H}_{\rm eff,u}$) and staggered field ($\mathbf{H}_{\rm eff,s}$) to an isotropic magnetic material.
Intuitively, it is then evident that in such a material the susceptibility toward a staggered order parameter is significantly enhanced, even up to large fields, in spite of the DM vector itself being comparatively small in magnitude.

This aspect was discussed in the context of analyzing the response of the QSL/VBS candidate $\kappa$-(BEDT-TTF)$_2$Cu$_2$(CN)$_3$ in the presence of an external magnetic field~\cite{Winter17importance} in $\mu$SR experiments~\cite{Pratt11}. However, while this ingredient is indeed important, 
a consistent interpretation of the numerous available experimental observations
seems to be only possible if disorder-induced orphan spins are also considered~\cite{Riedl19,Pustogow20,Miksch21,Matsuura22}. Hence, we will discuss the full description of this material in detail in Section~\ref{sec:disorder}.

\subsection{Four-Spin Ring Exchange}

As mentioned above, magnetic materials may be described with a pure spin model if they are in the strongly localized limit, i.e., $t/U \ll 1$. In this limit, it is  possible to extract, for instance, the bilinear Heisenberg model via second-order perturbation theory from the Hubbard model, where the Heisenberg exchange scales roughly with $t^2/U$.
Since the organic compounds considered in this review are close to the Mott MIT, these materials should, rather, be placed in the $t \lesssim U$ regime. When constructing a spin model for this class of materials, it may, therefore, be necessary to consider higher order contributions.
The most dominant higher order spin contribution is an isotropic four-spin ring exchange~\cite{motrunich2005}:
\begin{linenomath}
\begin{equation} \label{eq:ring-exchange_Hamiltonian}
    \mathcal{H}_{(4)} = \frac{1}{S^2} \sum_{\langle ijkl \rangle} K_{ijkl} \, [ (\mathbf{S}_i \cdot \mathbf{S}_j) (\mathbf{S}_k \cdot \mathbf{S}_l) 
    + (\mathbf{S}_i \cdot \mathbf{S}_l) (\mathbf{S}_j \cdot \mathbf{S}_k)
    - (\mathbf{S}_i \cdot \mathbf{S}_k) (\mathbf{S}_j \cdot \mathbf{S}_l) ],
\end{equation}
\end{linenomath}
defined on a plaquette spanned by the sites $\langle ijkl \rangle$. Note that we do not consider here four-spin terms arising from spin-orbit coupling effects. In Figure~\ref{fig:non-Heisenberg}a, the two distinct four-site plaquettes on the anisotropic triangular lattice are illustrated and labelled with $K$ and $K^\prime$. From perturbation theory, the four-site ring exchange scales with $t^4/U^3$, increasing in magnitude as $t$ approaches $U$. For certain ratios of $J^\prime/J$ and $K/J$, ring exchange can suppress magnetic order~\cite{Block11,Holt14} or induce new types of orders~\cite{cookmeyer2021}.
In Table~\ref{tab:exchange}, we list selected ab-initio results~\cite{Winter17importance, Riedl19, Riedl21, Riedl19thesis} for representative $\kappa$-phase salts. For all listed materials, the four-spin ring exchange is $\sim$5--10\% of the nearest-neighbor Heisenberg exchange. This is a comparatively strong contribution consistent with the proximity of the organic materials to the Mott MIT ($t \lesssim U$).
Note that it is not required by symmetry that the three terms in Equation~\eqref{eq:ring-exchange_Hamiltonian} have precisely identical prefactors. Since they are very similar for the considered materials, we discuss here averaged values for simplicity.

\begin{figure}[H]
\centering
\includegraphics[width=\columnwidth]{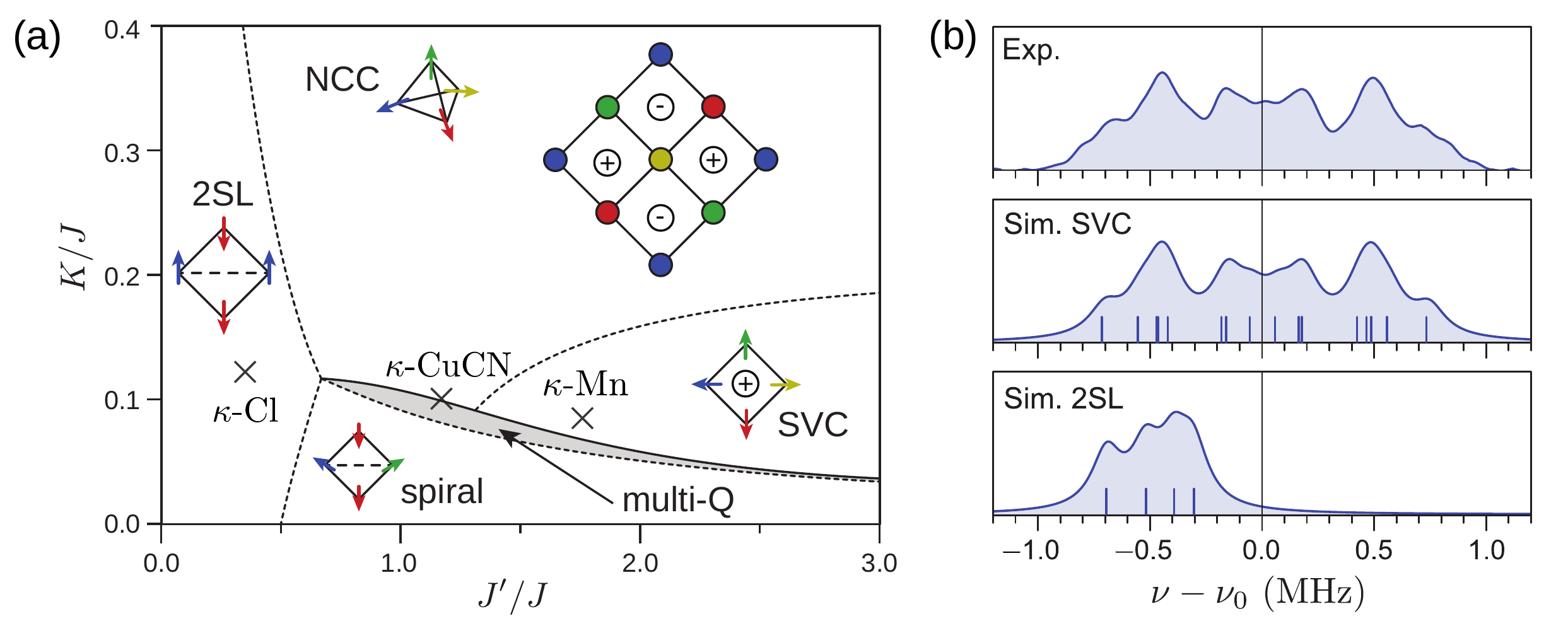} 
\vspace{-15pt}
\caption{(\textbf{a}) Classical phase diagram for the four-spin ring-exchange on the anisotropic triangular lattice. Indicated are the locations for $\kappa$-(BEDT-TTF)$_2$Cu[N(CN)$_2$]Cl ($\kappa$-Cl), $\kappa$-(BEDT-TTF)$_2$Cu$_2$(CN)$_3$ ($\kappa$-CuCN), and $\kappa$-(BETS)$_2$Mn[N(CN)$_2$]$_3$ ($\kappa$-Mn), based on ab-initio
calculations in Refs.~\cite{Winter17importance, Riedl19, Riedl21, Riedl19thesis} (see Table~\ref{tab:exchange}). The depicted phases include the two-sublattice (2SL) N\'eel order, non-coplanar chiral (NCC) order, and spin-vortex crystal (SVC) order. For this phase diagram, the constraints $J^\prime/J = J^{\prime\prime\prime}/J^{\prime\prime} = K^\prime/K$, $K/J^{\prime\prime} = 2$, and $|\mathbf{D}| = |\mathbf{\Gamma}| = 0$ were enforced. 
(\textbf{b}) $^{13}$C NMR spectra measured in Ref.~\cite{Vyaselev12} (``Exp.'') and simulated for hypothetical SVC and 2SL phases~\cite{Riedl21}.
Figures adapted from~Ref.~\cite{Riedl21}. }
\label{fig:ring-exchange}
\end{figure}

The corresponding classical phase diagram on the triangular lattice~\cite{Riedl21} is shown in Figure~\ref{fig:ring-exchange} using constraint ratios of the isotropic couplings suggested by perturbation theory~\cite{Holt14} and setting anisotropic couplings to zero.  
This phase diagram contains the well-known phases of the bilinear models on the anisotropic triangular lattice, the two-sublattice N\'eel order (``2SL'') and a spiral order with ordering wave vector $Q=(q,q)$, including the so-called ``120$^\circ$ order'' at the limit of an isotropic triangular lattice with $J^\prime/J=1$. In addition, a four-spin ring exchange induces novel phases, for instance four-sublattice orders, such as the non-coplanar chiral (NCC) order and the coplanar spin-vortex crystal (SVC) order. In the NCC phase, the magnetic moments are arranged such that, if they would be arranged at the corner of a tetrahedron, they would  point toward or away from the center of the tetrahedron. The sign of the magnetic moments on a four-site plaquette can be summarized by a sign of the plaquette, as indicated in the NCC phase in Figure~\ref{fig:ring-exchange}. The SVC phase is also a four-sublattice phase with vortex orientations, but in this case, the magnetic moments are constrained to a single plane.

Based on the ab-initio 
values in Table~\ref{tab:exchange}, three representative organic charge-transfer salts can be placed in the phase diagram in Figure~\ref{fig:ring-exchange}:
$\kappa$-(BEDT-TTF)$_2$Cu[N(CN)$_2$]Cl \mbox{($\kappa$-Cl)}, \mbox{$\kappa$-(BEDT-TTF)$_2$Cu$_2$(CN)$_3$} ($\kappa$-CuCN), and \mbox{$\kappa$-(BETS)$_2$Mn[N(CN)$_2$]$_3$} ($\kappa$-Mn).
Within the square lattice regime ($J^\prime/J \ll 1$), the four-spin ring exchange does not change the magnetic ground state. Consequently, while $\kappa$-Cl is found to have a significant ring-exchange contribution, its ground state is the two-sublattice AFM N\'eel state, consistent with experiments and as determined with the purely bilinear spin model.

In contrast, in the regime of an isotropic triangular lattice ($J^\prime/J \approx 1$), the four-spin ring exchange starts to play an important role for the magnetic ground state, so that \mbox{$\kappa$-CuCN} as significant ring exchange
is placed in the classical phase diagram close to phase boundaries between the multi-$Q$ spiral phase and the NCC phase. Quantum fluctuations may suppress the long-range order, entering a QSL/VBS phase, as suggested by semi-classical~\cite{Holt14} and DMRG~\cite{Block11} calculations. If disorder would be absent in $\kappa$-CuCN (see Section~\ref{sec:disorder}), the magnetic interactions
including four-spin ring exchange would point to a QSL/VBS ground state. 

A third consequence of a nonzero four-spin ring exchange is realized in $\kappa$-Mn~\cite{Riedl21}, where the higher order magnetic exchange does not suppress magnetic order, but instead selects an unconventional four-sublattice magnetic order, the SVC phase, which could not be accessed without a finite $K$. 
The SVC and NCC phases introduced by a significant ring exchange (see Figure~\ref{fig:ring-exchange}a) are characterized by a vector chiral order parameter $\vert \mathbf{v}_p \vert$, defined on a four-site square plaquette built by the solid bonds in Figure~\ref{fig:non-Heisenberg}a. The vector chirality $\mathbf{v}_p=\mathbf{S}_i\times\mathbf{S}_j + \mathbf{S}_j\times\mathbf{S}_k + \mathbf{S}_k\times\mathbf{S}_l + \mathbf{S}_l\times\mathbf{S}_i$ is finite when nearest-neighbor spins are orthogonal, but second-neighbor spins are antiparallel, as realized in the NCC and SVC phases. In the two-sublattice N\'eel order and spiral phases, the vector chirality vanishes: $\vert \mathbf{v}_p \vert=0$.
While it is difficult to estimate the influence of quantum fluctuations based on solely the classical phase diagram, there is a some experimental evidence for $\kappa$-Mn, which seems to require such a four-sublattice chiral phase. 
For $\kappa$-Mn, the interpretation of the magnetic properties from experimental evidence was initially challenged by the fact that the anion layer is composed of magnetic $S=5/2$ manganese atoms forming a distorted triangular lattice. However, the ordering temperature $T_{\rm N}=22\,$K~\cite{Vyaselev11,Vyaselev12,Vyaselev17} of $\kappa$-Mn can be assigned to the organic BETS spins, based on the fact that the ab-initio
exchange between Mn spins is with $\sim$1\,K two orders of magnitude smaller and that the entropy change around $T_{\rm N}$ observed in specific heat measurements is too small (8\% of $R\ln{2}$) for the significant participation of Mn spins~\cite{Riedl21}. For comparison, in the compounds $\lambda$-(BETS)$_2$FeCl$_4$ and $\kappa$-(BETS)$_2$FeBr$_4$, the Fe$^{3+}$ and organic spins order simultaneously~\cite{Mori02,Konoike04,Kartsovnik16}. Such features as Jaccarino--Peter superconductivity~\cite{jaccarino1962} and beats in the Shubnikov--de Haas effect, as reported in the latter compounds,
are not observed in $\kappa$-Mn~\cite{Zverev10,Vyaselev11,Zverev19}.
One example, where the experimental evidence of $\kappa$-Mn is not compatible with a 2SL or spiral order phase is the magnetic torque as a function of magnetic field~\cite{Vyaselev17}. 
As mentioned above, the out-of-plane and in-plane components of the DM vector differ in their pattern by symmetry. As a result, the vector chirality couples linearly only to $\mathbf{D}_b$, the contribution along the dashed bonds in Figure~\ref{fig:non-Heisenberg}a, confining the spins to lying in the $a^\ast c$ plane, while the 2SL and spiral phases couple only to $\mathbf{D}_{a c}$, confining the spins to the plane perpendicular to this contribution. In magnetic torque experiments, the features associated with BETS spins vanish for magnetic fields in the $a^\ast c$ plane, suggesting that the field couples to an order parameter for which this plane is a special plane of symmetry. This is only true for the NCC and SVC phases.
Additionally, $^{13}$C NMR experiments~\cite{Vyaselev12,Vyaselev12b} show signatures which are by symmetry not compatible with two-sublattice orders, but can be fitted well with the SVC order.
The experimental NMR resonance~\cite{Vyaselev12} (top panel in Figure~\ref{fig:ring-exchange}b) is symmetrical about the Larmor frequency $\nu_0$ with a rich fine structure. The expected resonance patterns were simulated in Ref.~\cite{Riedl21} using an ab-initio hyperfine coupling tensor, Lorentzian broadening and neglecting the Mn dipolar fields. As evident from Figure~\ref{fig:ring-exchange}b, the 2SL phase can be immediately ruled out, with only four distinct resonances, which are not symmetric around $\nu_0$. In contrast, the SVC phase shows 16 distinct, symmetric resonances, allowing for an excellent simulation of the experimental data.
Taking the classical phase diagram together with the experimental evidence on magnetic properties of $\kappa$-Mn, this BETS compound is a prime example for a material with the exotic magnetic order of the spin-vortex crystal phase, induced by the four-spin ring exchange.

\section{New Physics in the Extended Molecule-Based Model}
\label{sec:molecule-based}

In this section, we focus on experimental and theoretical observations that strongly suggest that a treatment of the $\kappa$-phase charge-transfer salts in the full molecule-based model is not only needed, but also can be the source of entirely new ordering phenomena. In particular, we
provide an in-depth review on the role of intra-dimer charge degrees of freedom in generating electronically driven ferroelectric ground states and
we  discuss the role of the anisotropy of the three inter-dimer hoppings that exist in the molecule model for the pairing symmetry of unconventional superconductivity.

\subsection{Ferroelectricity in the Mott Insulating Ground State}
\label{sec:intra-dimer}

When changing from the effective half-filled one-band dimer model on the triangular lattice  with 
on-site Coulomb repulsion $U$ per dimer and hopping terms $t$, $t^\prime$ between dimers
to the molecule-based model, the following energy scales are at place: the hopping parameters $t_{1\ldots 4}$ as shown in Figure~\ref{fig:overview}, 
a Coulomb repulsion $\tilde{U}$ on the molecule (please note that $U$ was defined as the onsite Coulomb repulsion on the dimer, while  $\tilde{U}$ corresponds
to the onsite Coulomb repulsion on the molecule), and additional inter-molecule Coulomb repulsion terms $V$,
which are generally expected to be smaller than $\tilde{U}$ (typically \mbox{$V/\tilde{U}\,\sim\,$0.4)~\cite{Hotta10,Kaneko17}.} 
 Including $V$ in the considerations introduces the possibility of intra-dimer charge orders that may compete with the dimer-Mott insulating state \cite{Seo04,Hotta10,Kaneko17}. As a result, a new degree of freedom, namely, intra-dimer charge fluctuations, may emerge in dimerized $\kappa$-phase charge-transfer salts \cite{Hotta10,Kaneko17}.

Many of the theoretical considerations regarding the inclusion of intra-dimer charge degrees of freedom into minimal low-energy models were further motivated by the simultaneous experimental discoveries of dielectric anomalies in various Mott insulating $\kappa$-phase charge-transfer salts \cite{AbdelJawad10,Lunkenheimer12,Gati18,Lunkenheimer15b}. There is a large body of evidence suggesting that these anomalies result from ferroelectricity which might be driven by the active intra-dimer charge degrees of freedom. 

\begin{figure}[t]
\centering
\includegraphics[width=0.98\textwidth]{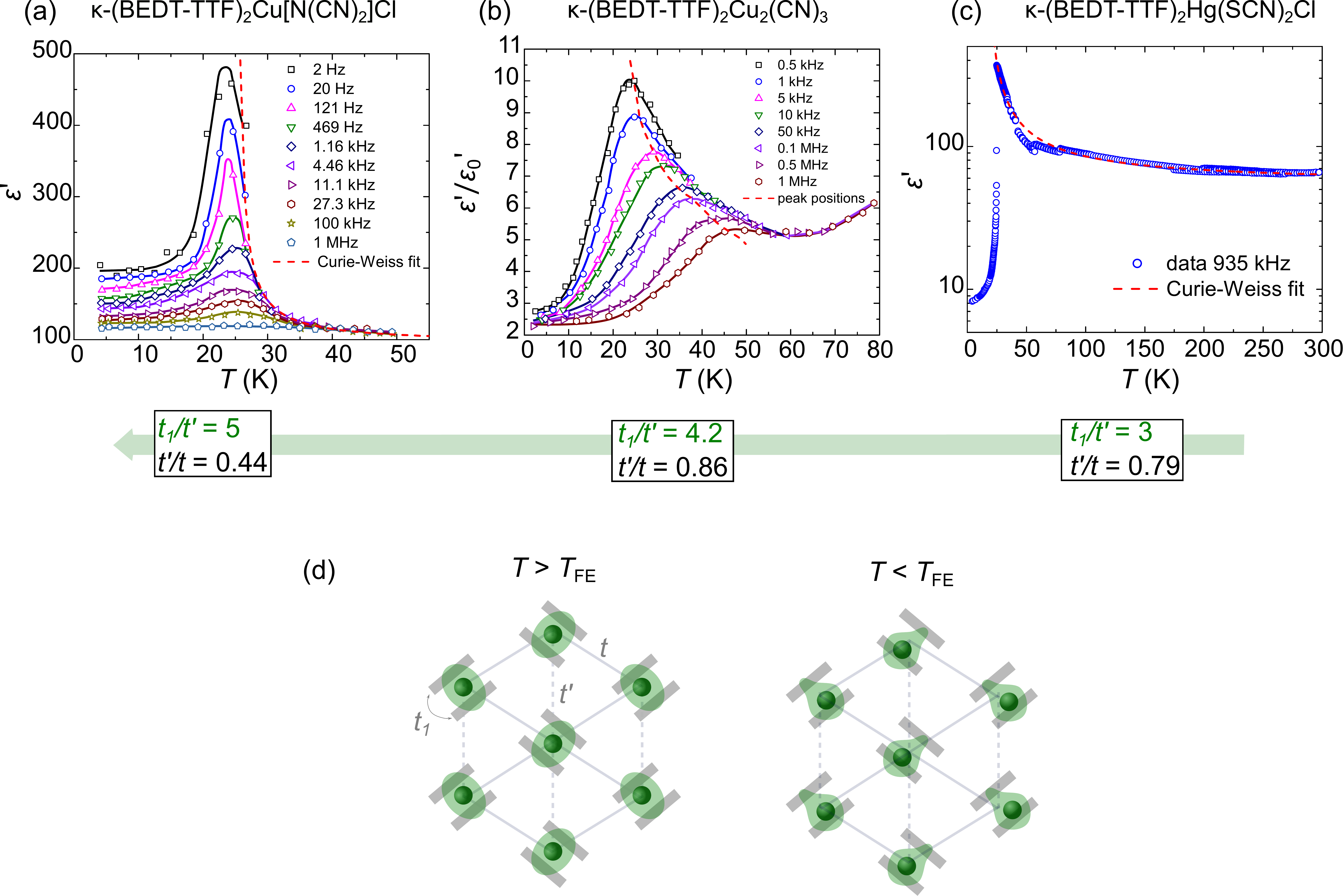} 
\caption{Role of intra-dimer charge degrees of freedom revealed by dielectric measurements: \linebreak \mbox{(\textbf{a}--\textbf{c})} Frequency-dependent dielectric constant $\epsilon^\prime$ on three $\kappa$-phase organic charge-transfer salts: \linebreak (\textbf{b}) $\kappa$-(BEDT-TTF)$_2$Cu[N(CN)$_2$]Cl ($\kappa$-Cl)~\cite{Lunkenheimer12,Lang14,Lunkenheimer15b}, (\textbf{c}) $\kappa$-(BEDT-TTF)$_2$Cu$_2$(CN)$_3$ ($\kappa$-CuCN)~\cite{AbdelJawad10} and (\textbf{d}) $\kappa$-(BEDT-TTF)$_2$Hg(SCN)$_2$Cl~\cite{Gati18}. Symbols represent the measured data, solid lines are guide to the eyes. The red dashed lines represent Curie--Weiss fits of the data in (\textbf{b},\textbf{d}), whereas they show the frequency dependence of the peak position in (\textbf{c}). These three materials differ by the degree of dimerization ($t_1/t^\prime$), as well as the frustration strength ($t^\prime/t$, defined in the effective dimer model). Values for $t_1/t^\prime$ and $t^\prime/t$ for each of the materials are included below the figures and taken from Refs.~\cite{Kandpal09,Jeschke12,Gati18}; (\textbf{d}) Schematic view of charge distribution on the dimerized $\kappa$-phase structure. For temperatures above the ferroelectric transition temperature (i.e., $T$\,>\,$T_{\rm FE}$), charge is equally distributed within a dimer. For $T$\,<\,$T_{\rm FE}$, intra-dimer charge order sets in, giving rise to a macroscopic polarization.}
\label{fig:dielectricfigure}
\end{figure}

Dielectric spectroscopy is a common tool that is sensitive to ferroelectric order \cite{Lunkenheimer15b}. A typical signature for a ferroelectric transition is a peak in the temperature dependence of the dielectric constant $\epsilon^\prime(T)$. Such a peak was indeed observed for various $\kappa$-phase charge-transfer salts in the Mott insulating state, as depicted in Figure\,\ref{fig:dielectricfigure}. First, it was discovered in $\kappa$-(BEDT-TTF)$_2$Cu$_2$(CN)$_3$ ($\kappa$-CuCN, Figure\,\ref{fig:dielectricfigure}b) below $T$ $\lesssim$ 60\,K by Abdel-Jawad et al. \cite{AbdelJawad10}. Here, the peak of the dielectric constant shifts with frequency, which is a characteristic for a ``relaxor ferroelectric''. In line with this classification, the peak temperatures as a function of frequency are well-described by a Vogel--Fulcher--Tammann equation with characteristic temperature of $T_{\rm VFT}$\,$\sim$\,6\,K, which interestingly corresponds to the temperature of the famous ``6 K anomaly'' in this compound \cite{Manna10,Miksch21}. A relaxor ferroelectric does not manifest long-range order, but rather represents a cluster-like order mediated by short-range correlations.

In contrast to this, a so-called ``order--disorder''-type transition, where 
the peak position is independent of frequency, was found in \mbox{$\kappa$-(BEDT-TTF)$_2$ Cu[N(CN)$_2$]Cl} ($\kappa$-Cl) \cite{Lunkenheimer12,Lang14}, as shown in Figure\,\ref{fig:dielectricfigure}a.
This notion was further corroborated by measurements of the ferroelectric hysteresis as well as measurements of the switchability of the ferroelectric polarization. The dielectric constant in $\kappa$-Cl at high temperatures is well-described with a Curie--Weiss law with a characteristic temperature $T_{\rm CW}$\,$\sim$\,27\,K. This temperature is identical to the antiferromagnetic ordering temperature $T_{\rm N}$ of this compound, making this material a realization of a multiferroic system \cite{Lunkenheimer12}---an intriguing state of matter with cross-coupling between charge and spin degrees of freedom. Independent of the detailed behavior as a function of temperature or frequency, it can be concluded that these dielectric measurements indicate active charge degrees of freedom deep inside the Mott insulating state, which casts doubts on the stringent applicability of the dimer model.

In fact, intra-dimer charge order would be a natural explanation for the formation of ferroelectricity in the $\kappa$-phase charge-transfer salts. In general, ferroelectricity requires (i)~the existence of dipoles and (ii) the breaking of inversion symmetry for the formation of a macroscopic polarization. In case of intra-dimer charge order, both conditions are naturally fulfilled in the dimerized $\kappa$-phase structure, see Figure\,\ref{fig:dielectricfigure}d. At high temperatures above the ferroelectric transition temperature $T$\,>\,$T_{\rm FE}$, the charges are localized on a dimer, corresponding to fluctuating dipoles \cite{Itoh13}. When cooling below $T_{\rm FE}$, the charges localize on one of the two organic molecules that form a dimer. As a result, a static dipole is created. At the same time, through localization on one of the molecules, inversion symmetry is broken. Thus, overall a macroscopic polarization arises. This electronically driven mechanism for ferroelectricity, which is invoked for both salts $\kappa$-Cl and $\kappa$-CuCN \cite{AbdelJawad10,Lunkenheimer12,Hotta10,Gomi10,Gomi13}, is different from the off-center displacement of ions in conventional ferroelectrics \cite{vandenBrink08}.

Clearly, the electronically driven scenario requires a proof of intra-dimer charge order, similar to the case of weakly dimerized quasi one-dimensional TMTTF-based~\cite{Nad06} as well as $\alpha$-phase BEDT-TTF salts \cite{Lunkenheimer15,Takano01}, where charge order and ferroelectricity are well established. However, attempts to resolve a charge disproportionation within the BEDT-TTF dimer for \mbox{$\kappa$-Cl} and $\kappa$-CuCN have remained unsuccessful \cite{Tomic15}. By monitoring charge-sensitive phonon vibrational modes in infrared spectroscopy, no indications for a charge-order-induced splitting of modes were observed for $\kappa$-Cl and $\kappa$-CuCN within their experimental resolution~\cite{Sedlmeier12}. This puts an upper limit on the possible size of charge disproportionation of $\delta$\,$\sim$\,$\pm\,0.005$\,e. The absence of a detectable charge disproportionation has motivated proposals of alternative explanations for the observation of dielectric anomalies~\cite{Tomic13,Pinteric14,Pinteric16,Pustogow21}. Prominent proposals include (i) charged domain-wall relaxations in the weak ferromagnetic state at lower temperatures for $\kappa$-Cl \cite{Tomic13}, (ii) charge defects triggered by a local inversion-symmetry breaking in the anion in $\kappa$-CuCN~\cite{Pinteric14,Pinteric16,Dressel16,Lazic18} and (iii)~a percolative enhancement of the dielectric constant in proximity to a Mott MIT based on experiments on $\kappa$-CuCN~\mbox{\cite{Pustogow21,Roesslhuber21}}. Besides controversies regarding the data analysis of the dielectric data, see, e.g., Refs.~\mbox{\cite{Lang14,Lunkenheimer15b,Tomic15}}, there is another concern that some of the alternative explanations for the dielectric anomalies involve material-specific arguments. This makes these proposals a possible explanation for the respective specific compounds, but does not answer the question of why so many $\kappa$-phase charge-transfer salts exhibit dielectric anomalies deep in the Mott insulating state. We note that the proposal that the dielectric signal even deep in the Mott insulating state arises from the spatial coexistence of insulating and correlated metallic phases close to the Mott MIT~\cite{Pustogow21,Roesslhuber21} does not necessarily explain the emergence of long-range order-disorder-type ferroelectricity.

Within the picture of intra-dimer charge order, the degree of charge disproportionation $\delta$ is expected to relate to the strength of dimerization. From ab-initio model parameters, $\kappa$-Cl and $\kappa$-CuCN both fall in the same range of rather strong dimerization with \mbox{$t_1/t^\prime$\,$\sim$\,4.2--5 \cite{Kandpal09,Jeschke12}} (see Figure\,\ref{fig:dielectricfigure}). Thus, a clearer case for electronically driven ferroelectricity might be made when dimerization is still dominant, but weaker than in the materials cited above. Indeed, according to DFT calculations, the related material $\kappa$-(BEDT-TTF)$_2$Hg(SCN)$_2$Cl is characterized by such a moderate degree of dimerization \mbox{$t_1/t^\prime$\,$\sim$\,3 \cite{Gati18}}. For this compound, charge order with $\delta\,=\,\pm\,0.1$ e was unequivocally identified in vibrational spectroscopy \cite{Drichko14}. Dielectric data for this compound \cite{Gati18} is shown in Figure\,\ref{fig:dielectricfigure}c. The behavior of $\epsilon^\prime(T)$ is very reminiscent of the behavior of the characteristics of well-established order-disorder type ferroelectrics \cite{Lines77}. Whereas in $\kappa$-(BEDT-TTF)$_2$Hg(SCN)$_2$Cl there is, thus, strong evidence for long-range charge~\cite{Drichko14} and ferroelectric order~\cite{Gati18}, there is no long-range charge order in its sister compound $\kappa$-(BEDT-TTF)$_2$Hg(SCN)$_2$Br. Nonetheless, experimental evidence from Raman spectroscopy for a ``quantum dipole liquid'' was presented for the Br-variant, in which electric dipoles from intra-dimer charge degrees of freedom remain fluctuating down to lowest temperatures~\cite{Hassan18}. Overall, the collection of data on both compounds \mbox{$\kappa$-(BEDT-TTF)$_2$Hg(SCN)$_2X$} with $X$\,=\,Cl, Br \cite{Gati18,Hassan18,Hassan20} provide strong evidence that intra-dimer charge degrees of freedom are relevant in dimerized $\kappa$-phase materials.

\subsection{Superconductivity in Extended Molecule-Based Models}

There are also distinct differences between the physics of the dimer model and the extended molecule-based model in the metallic regime as a result of (i) intra-dimer Coulomb interactions and (ii) the anisotropy of the inter-dimer hoppings. The relevance of the latter becomes evident in the discussion of the superconducting order parameter of the $\kappa$-phase materials. In fact, following previous work by Kuroki {et al.}~\cite{Kuroki02}, Guterding {et al.}~\cite{Guterding16b,Guterding16} calculated the superconducting pairing symmetry of a few $\kappa$-based charge-transfer-salts superconductors using a random phase approximation (RPA) spin-fluctuation approach and hopping parameters extracted from ab-initio-based density functional theory calculations in the molecule description. It was found that the order parameter is substantially altered when considering the anisotropy in $t_2$ and $t_4$ (defined in Figure~\ref{fig:overview}), which is averaged out in the dimer model. Sufficiently large anisotropy promotes a peculiar competition between square-like and diagonal hoppings, which is absent in the effective dimer model. As a result, for significant anisotropy, mixed-order parameters of type $s_{\pm}+d_{x^2-y^2}$ are favored over the single-component $d_{xy}$ (in the notation of the physical Brillouin zone). Qualitatively, these results were later reinforced by solving the linearized Eliashberg equation using the two-particle self-consistent approach \cite{Zantout18}, even though there is a quantitative difference in the precise location of the crossover from $d_{xy}$ to $s_{\pm}+d_{x^2-y^2}$. 

\begin{figure}[H]
\centering
\includegraphics[width=\textwidth]{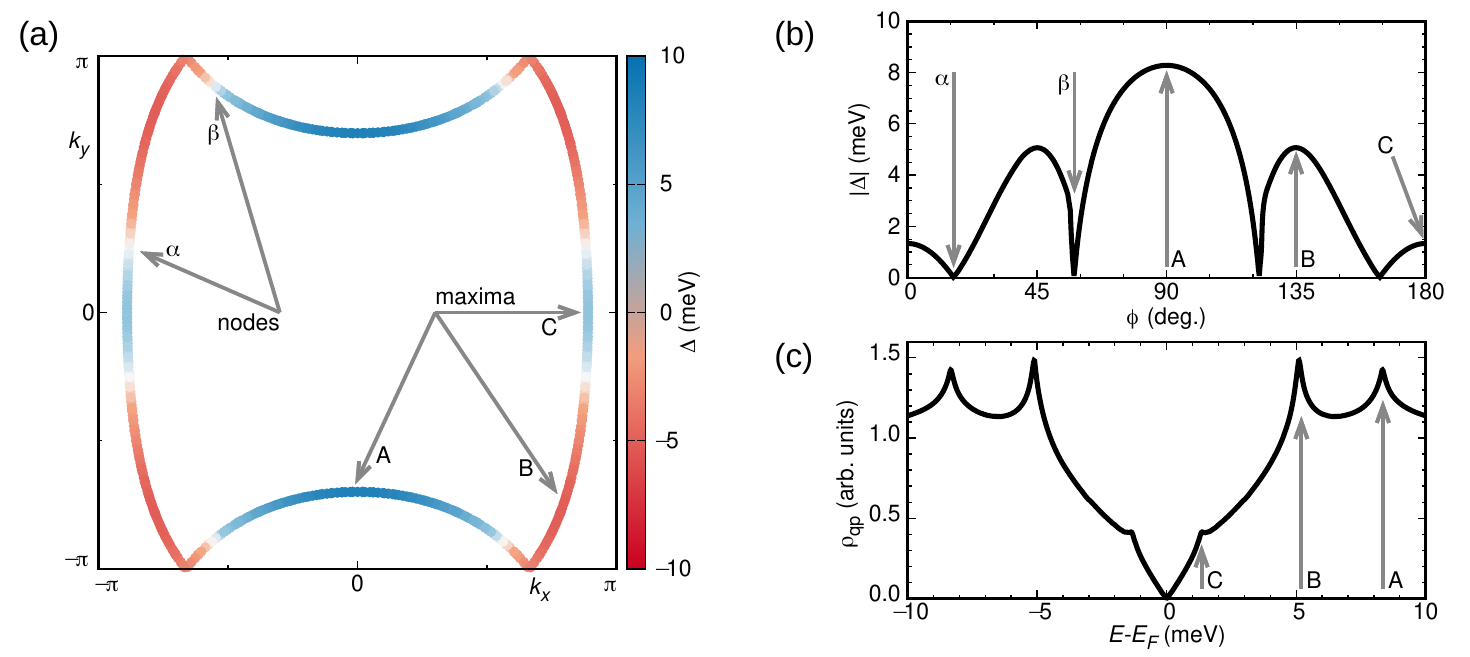} 
\vspace{-6pt} 
\caption{Theoretical results for $\kappa$-(BEDT-TTF)$_2$Cu[N(CN)$_2$]Br, calculated by an RPA spin-fluctuation approach using  hopping parameters extracted from ab-initio-based density functional theory
calculations in the extended molecular model: (\textbf{a}) Eight-node mixed-symmetry superconducting gap $\Delta$, (\textbf{b})~$| \Delta|$ as a function of the angle $\phi$ with respect to $k_x$, (\textbf{c}) quasi-particle DOS $\rho_{\rm qp}$ in the superconducting state. The three coherence peaks A, B, C are consistent with scanning tunneling spectroscopy results~\cite{Guterding16}. Figures adapted from Ref.~\cite{Guterding16}.
}
\label{fig:SC-gap}
\end{figure}

The theoretically predicted mixed-order parameter in the full molecule-based model has eight nodes (see Figure~\ref{fig:SC-gap}a), giving rise to three distinct coherence peaks (Figure~\ref{fig:SC-gap}b,c). The existence of the three coherence peaks was confirmed in low-temperature scanning tunneling spectroscopy of  $\kappa$-(BEDT-TTF)$_2$Cu[N(CN)$_2$]Br~\cite{Guterding16}. It is also interesting to note that calculations on the extended Hubbard model using Monte Carlo simulations found that superconductivity with mixed $s+d$ order parameter is stabilized on the verge of charge-order and dimer-Mott instabilities \cite{Watanabe17,Sekine13,Gomes15}.

\vspace{-6pt} 

\section{Coupling of Correlated Electrons to the Crystal Lattice}
\label{sec:lattice}

Treating the organic $\kappa$-phase charge-transfer salts or any other correlated electron system in terms of a purely electronic Hubbard model does not consider that charge and spin degrees of freedom are coupled to the crystal lattice \cite{Majumdar94,Hassan05}. That this electron-lattice coupling could have significant consequences for the behavior of the correlated electron system might already be inferred from the fact that physical pressure is often used to induce phase transitions and associated critical behaviors. For example, as outlined in Section~\ref{sec:phase-diagram}, pressure can be used to induce the Mott MIT. Importantly, the underlying compressibility of the crystal lattice does also necessarily imply that the crystal lattice can respond to the correlated electron system in non-trivial ways. Since such a type of coupling exists in any solid-state realization of a correlated electron system, understanding the consequences of the electron--lattice coupling is nowadays central to various fields of research. A prominent example is the electronic nematic phase which is believed to be ubiquitous to the phase diagrams of many high-temperature superconductors \cite{Kuo16,Fernandes14,Achkar16} and which is intimately coupled to lattice instabilities~\cite{Cano10,Boehmer16,Gati20}. In addition, the coupling to the lattice has received increasing levels of attention in the field of organic \cite{Manna10,Miksch21,Manna14,Manna18,Hartmann18}  and inorganic frustrated magnets \cite{Sushkov17,Li20,Kaib21,Ferrari20,Ferrari21,Brenig20,Ye20}.

In the following, we deepen the discussion of the underlying ideas by analyzing recent experimental results for the $\kappa$-phase organic salts~\cite{Gati17,Matsuura19}. The high-pressure sensitivity of these materials suggests that they are characterized by a comparatively large coupling of the electrons to the crystal lattice~\cite{Mueller02}. In this sense, the $\kappa$-phase 
charge-transfer salts are an ideal testground to discover and understand novel effects that arise from electron--lattice coupling, since any effect will be significantly amplified when such a coupling is naturally~strong.

\subsection{Critical Elasticity around the Mott Critical Endpoint}

We start by a discussion of how coupling of correlated electrons to the crystal lattice impacts the pressure-induced Mott transition.  For that, we review theoretical ideas and experimental results on $\kappa$-(BEDT-TTF)$_2$Cu[N(CN)$_2$]Cl ($\kappa$-Cl). The purely electronic Mott critical endpoint is characterized by a diverging response function \cite{Kotliar00,Castellani79}. When such an electronic system is embedded into a crystal lattice with finite compressibility, then the lattice responds to changes in the electronic system \cite{Majumdar94,Hassan05,Zacharias12}. Sufficiently away from the critical endpoint, the crystal lattice will barely respond to the only weakly fluctuating electronic degrees of freedom. However, close to the endpoint, when the electronic system becomes critical, the internal pressure exerted by the electronic system will induce additional lattice strain. In other words, the crystal lattice is driven soft, leading to a renormalization of the system's compressibility due to a vanishing elastic modulus. These renormalization effects can successfully be captured in DMFT calculations of the compressible Hubbard model \cite{Majumdar94,Hassan05}.

Importantly, the softening of the crystal lattice implies that the crystal lattice itself becomes critical when the electronic degrees of freedom become critical. This was pointed out in Ref.\,\cite{Zacharias12} based on considerations of an effective field theoretical description of the coupling of the Mott order parameter to lattice strain. The electronic Mott transition is, hence, expected to be preempted by an isostructural transition. Consequently, the criticality of the Mott system coupled to lattice degrees of freedom is described by mean field criticality, the universality class of isostructural solid--solid transitions, where the shear forces of the crystal lattice strongly suppress fluctuations~\cite{Zacharias12,Zacharias15}.   

\begin{figure}[H]
\centering
\includegraphics[width=1\columnwidth]{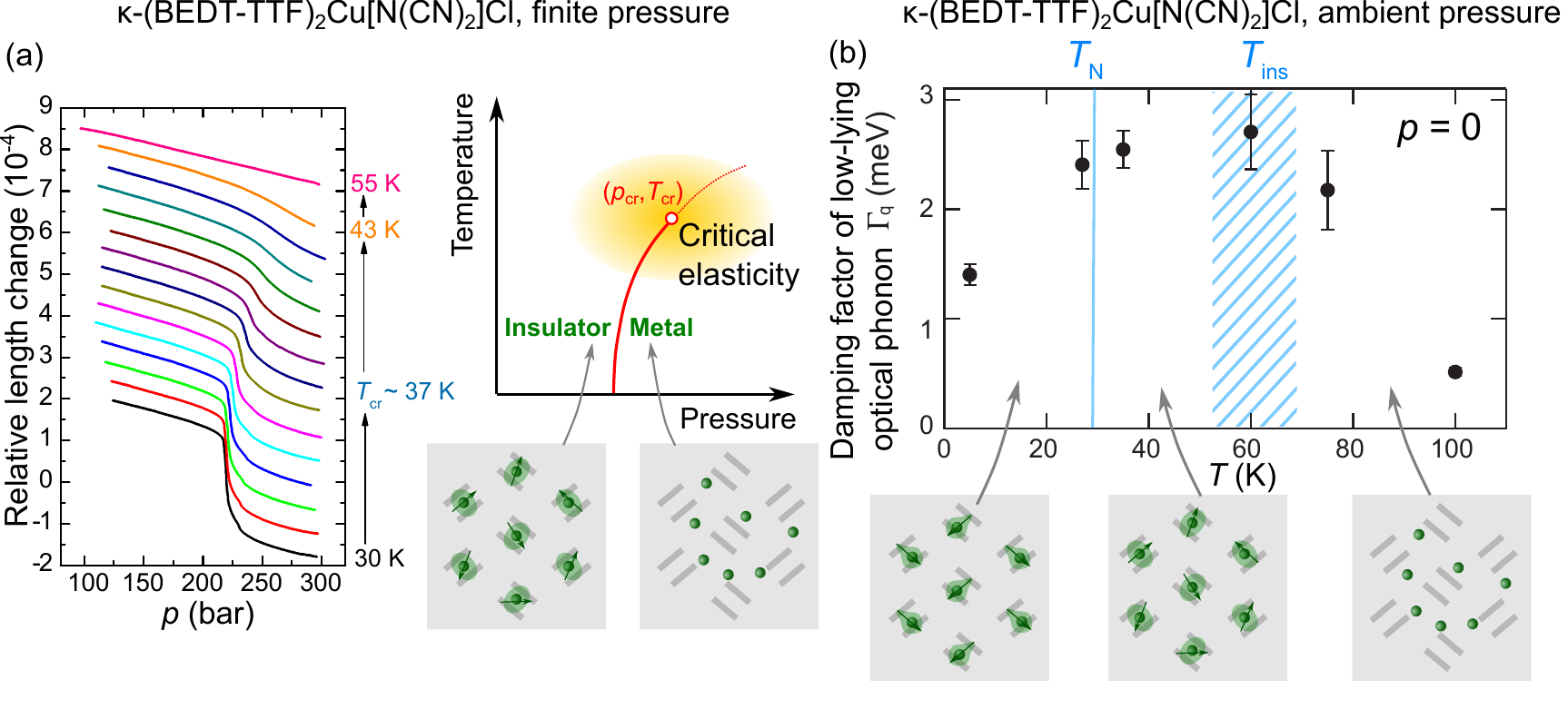} 
\caption{Coupling of correlated $\pi$-electrons with the crystal lattice in $\kappa$-(BEDT-TTF)$_2$Cu[N(CN)$_2$]Cl. (\textbf{a}) Strain--stress relationships across the pressure-induced Mott metal--insulator transition \cite{Gati17}: The relative length-change data on the left as a function of hydrostatic pressure indicate a wide temperature region above the critical endpoint located at $T_{\rm cr}$\,$\sim$\,37\,K, in which strongly non-linear strain--stress relationships are observed. This result implies that the crystal lattice itself becomes soft as the electronic degrees of freedom become critical. This effect has been termed ``critical elasticity'' \cite{Zacharias12,Zacharias15} and prevails in an extended region around the endpoint (right panel); (\textbf{b}) ambient-pressure study of the damping factor of a low-lying optical phonon, likely related to the BEDT-TTF breathing mode, as a function of temperature \cite{Matsuura19}. The damping factor is inversely proportional to the phonon lifetime. Three characteristic regimes were identified in the data: (i) low damping at high temperatures ($T\,>\,T_{\rm ins}$) before the charge gap opens, (ii) high damping when the charges localize on the dimer ($T_{\rm N}\,<\,T\,<\,T_{\rm ins}$) and (iii) low damping below the onset of spin and charge order ($T\,<\,T_{\rm N}$). Figures adapted from Refs.~\cite{Gati17,Matsuura19}.}
\label{fig:lattice-effects}
\end{figure}

Experimentally, the relevance of the theoretical considerations above was explicitly demonstrated by studies of the lattice strains as a function of pressure, i.e., strain--stress relations, around the Mott critical endpoint in $\kappa$-Cl in Ref.~\cite{Gati17}. The experimental data is shown in Figure\,\ref{fig:lattice-effects}a. The key result here is the observation of pronounced non-linearities in the strain--stress relationships at temperatures higher than the critical temperature \mbox{$T_{\rm cr}\,\approx\,37$\,K}. Even at a temperature of 43\,K, i.e., $(T-T_{\rm cr})/T_{\rm cr}\,\sim\,20\%$, the renormalization of the compressibility is found to be of the order of the compressibility itself. Thus, these results indicate a breakdown of Hooke's law induced by critical electronic degrees of freedom. The measured data do not only qualitatively agree with the physical picture described above, but also quantitatively. Following the classification as an isostructural solid--solid endpoint~\cite{Zacharias12}, for which the volume reflects an appropriate order parameter, the experimental data was quantitatively well described by the associated mean-field criticality \cite{Gati17}.

Taken together, theory and experiment suggest that any Mott system that is coupled to a compressible lattice will eventually be controlled by this ``critical elasticity''. Importantly, this effect can be dominant in a wide range of the temperature--pressure phase diagram, as experimentally proven for $\kappa$-Cl and schematically depicted on the right side of \mbox{Figure\,\ref{fig:lattice-effects}a~\cite{Gati17}}. Thus, the example of the Mott MIT and its criticality demonstrates nicely that electron--lattice coupling is not only a small, perturbative correction to the correlated electron problem, but an essential ingredient.

\subsection{Phonon Anomalies Probed by Inelastic Neutron Scattering}

The fact that the (de-)localization of the $\pi$ electrons at the Mott MIT causes drastic anomalies in the lattice properties is a very good indication that, vice versa, anomalies in the lattice dynamics can be a very sensitive probe of underlying fluctuating electronic degrees of freedom. Thus, probing lattice dynamics can be a very important tool to detect active charge and/or spin degrees of freedom, e.g., in the Mott insulating state \textls[-5]{where the role of intra-dimer charge degrees of freedom is particularly debated (see Section\,\ref{sec:intra-dimer},~above).}

Due to the large number of atoms per unit cell, there are typically many lattice and molecular vibration modes observed in, for instance, Raman or vibrational spectroscopy over a large energy range \cite{Dressel16}. The modes of interest that are expected to be particularly sensitive to fluctuating electronic degrees of freedom are energetically low-lying phonons, in particular those that involve breathing motion within the BEDT-TTF dimer. Based on ab-initio phonon calculations, those phonons typically lie in an energy range of just a few meV, e.g., at 4.1 and 4.7\,meV in $\kappa$-(BEDT-TTF)$_2$Cu$_2$(CN)$_3$ ($\kappa$-CuCN) \cite{Dressel16}. Such low-energy phonons can be probed with high sensitivity in neutron-scattering measurements. For a long time, neutron-scattering measurements on the organics have been rare due to the typically small size of the single crystals \cite{Pintschovius97}. However, technical improvements have now allowed to perform inelastic neutron-scattering measurements to probe phonon changes in $\kappa$-Cl \cite{Matsuura19} at ambient pressure and even more recently on $\kappa$-CuCN \cite{Matsuura22}.

The main result of the work on $\kappa$-Cl, as shown in Figure\,\ref{fig:lattice-effects}b, is that the damping of a low-lying optical phonon, located around 2.6\,meV, shows a strongly non-monotonic behavior as a function of temperature \cite{Matsuura19}. In particular, the phonon damping is specially high in an intermediate temperature range: When cooling $\kappa$-Cl from 100\,K to below \mbox{$\sim$50--60\,K}, the phonon damping increases rapidly, corresponding to a significantly reduced phonon lifetime. Upon further cooling below $T_{\rm N}$\,$\sim$\,27\,K, the phonon lifetime is found to be significantly enhanced again. Interestingly, the onset of phonon damping at $\sim$50--60\,K coincides with the opening of the charge gap at $T_{\rm ins}$ \cite{Sasaki04}. Below this temperature, the hole carriers become localized on the dimer, but the intra-dimer charge degree of freedom remains active. Only when the intra-dimer charge and the spin degrees of freedom are ordered below $T_{\rm N}$, is a truly inelastic behavior of the phonon modes recovered (see schematics in Figure\,\ref{fig:lattice-effects}b). While this result suggests an intimate coupling of the lattice with charge and spin degrees of freedom, which makes the lattice unstable, it is important to note that the neutron data does not support a structurally driven damping of the phonon modes. Thus, electronic degrees of freedom (charge and spin) are the driving force behind these lattice anomalies~\cite{Matsuura19}. 

A possible way to describe such a coupling is given by the pseudospin-phonon-coupled model, as proposed in Ref. \cite{Yamada74}, which can be extended to the coupling of phonons to other stochastic variables than only spins. According to this model, the characteristic energy of the electronic degree of freedom must be of the order of the phonon energy itself for significant phonon damping to occur. Interestingly, the intra-dimer charges that give rise to the dipole liquid in the related $\kappa$-(BEDT-TTF)$_2$Hg(SCN)$_2$Br \cite{Hassan18} (see Section\,\ref{sec:intra-dimer}) were found to be located at $\sim$40\,cm$^{-1}$, which is similar in energy to the overdamped optical phonon in $\kappa$-Cl. Thus, it seems likely that the overdamped phonons observed in inelastic neutron-scattering experiments in $\kappa$-Cl are related to fluctuating intra-dimer charge and spin degrees of freedom.

We note that the recent study of phonons in $\kappa$-CuCN \cite{Matsuura22} revealed strong similarities in the behavior of the phonon modes. This includes (i) a strongly overdamped phonon mode, when the $\pi$-electrons are localized on the dimers and (ii) a recovery of a long-lived phonon mode below the characteristic temperature of 6\,K. This result clearly strengthens the notion that the enigmatic ``6\,K anomaly'' is a result of a phase transition rather than a crossover. It was argued based on a spin-charge coupling model that the observed anomalous phonon behavior is consistent with a transition into a valence-bond solid below 6\,K.

\section{Role of Disorder}
\label{sec:disorder}

Disorder of various kinds is inevitably present in any real material. Even though observations of quantum oscillations in $\kappa$-phase metals in low fields and/or relatively high temperatures indicate the high crystalline quality of pristine samples \cite{Wosnitza93}, there is by now some body of evidence that disorder is important for understanding the properties of the $\kappa$-phase charge-transfer salts. In the present section, we discuss experimental results of the impact of controlled disorder on the properties close to the Mott MIT, as well as theoretical results on the role of disorder and resulting orphan spins in the magnetic phase.

\subsection{Experimental Study of Phenomena Close to the Mott Transition under Controlled Disorder}

\begin{figure}[H]
\centering
\includegraphics[width=0.9\columnwidth]{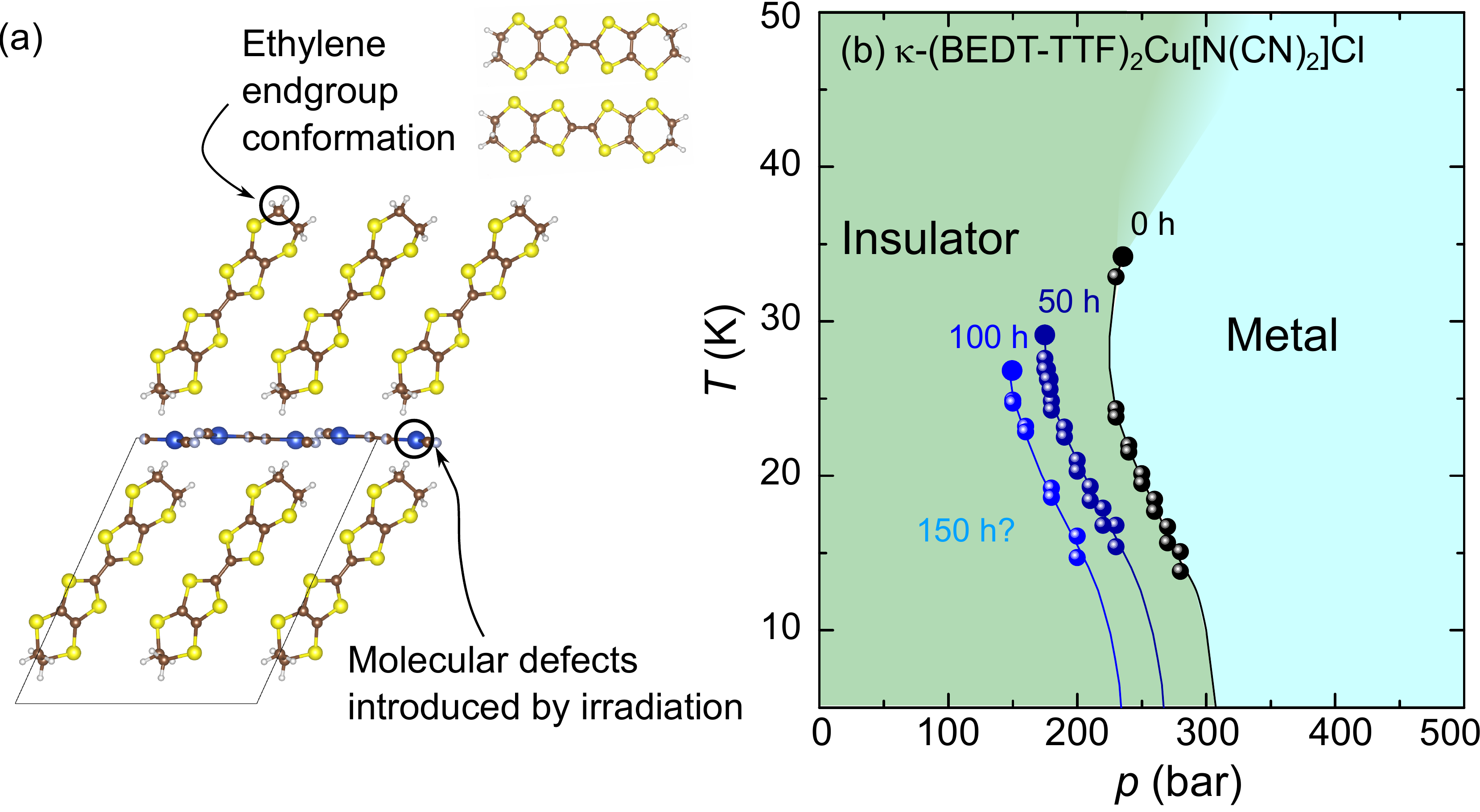} 
\caption{Experimental studies of the impact of weak disorder in $\kappa$-phase charge-transfer salts. (\textbf{a})~Experimentally, disorder can be deliberately introduced either by (i) creating molecular defects through X-ray irradiation \cite{Sasaki12} or (ii) by controlling the ethylene endgroup conformation via the cooling rate through the glass-forming temperature $T_{\rm glass}$ \cite{Mueller15}; (\textbf{b}) temperature--pressure phase diagram for $\kappa$-(BEDT-TTF)$_2$Cu[N(CN)$_2$]Cl, subjected to different irradiation times \cite{Gati18b}. Symbols represent the discontinuous Mott metal--insulator transition.
(\textbf{b}) adapted from Ref.~\cite{Gati18b}.}
\label{fig:shift-Mottline-disorder}
\end{figure}

To better grasp the effect of disorder, it is pivotal to develop means to controllably change the degree of disorder. For the $\kappa$-phase organic charge-transfer salts, two means are established with which disorder can be controlled, both reversibly and irreversibly (see Figure~\ref{fig:shift-Mottline-disorder}a): (i) control of the conformational degree of freedom of the ethylene endgroups (EEG) in the BEDT-TTF molecule by varying the cooling rate through the associated glass transition at $T_{\rm g}$ \cite{Mueller15}, and (ii) introduction of molecular defects, dominantly in the anion layer \cite{Kang17}, through X-ray irradiation \cite{Sasaki12}. In the remainder of this section, we summarize recent results, based on both techniques.

First, we focus on the reversible approach of introducing disorder through controlling the EEG disorder. Those ethylene endgroups ([C$_2$H$_4$]) can adopt two different configurations, when viewed along the central C$=$C bond. The EEGs can either show an eclipsed or a staggered configuration, see Figure\,\ref{fig:shift-Mottline-disorder}a. Upon cooling, the EEG tend to adopt one of the conformations. However, the ordering usually cannot be completed for kinetic reasons~\cite{Mueller15,Mueller02}. The associated relaxation becomes so slow close to $T_{\rm g}$ that equilibrium cannot be achieved, resulting in a certain amount of intrinsic structural disorder. Importantly, the amount of disorder can be controlled in a reversible manner by heating above $T_{\rm g}$, thus melting the frozen EEG configuration, and consecutively adjusting the cooling rate through the glass transition at $T_{\rm g}$.

An important task is to quantify the amount of disorder induced by the glassy EEG freezing. This task was addressed in Ref.\,\cite{Hartmann14} utilizing a special heat-pulse protocol to achieve large cooling rates up to several 1000\,K/min, together with modelling of the glass transition in terms of a double-well potential with realistic energy parameters \cite{Mueller15,Hartmann14,Gati18c}. It was found that EEG disorder levels of up to $\sim$6\% might be reached in real experiments with the largest cooling speeds. Conversely, this implies for the pristine, slowly cooled crystal that intrinsic EEG disorder levels are at a maximum about 2 to 3\%.

Experimentally, an increase in the residual resistivity, as well as an influence on the superconducting $T_{\rm c}$, were observed in metallic $\kappa$-(BEDT-TTF)$_2$Cu[N(CN)$_2$]Br upon cooling with higher speed through $T_{\rm g}$ \cite{Su98,Yoneyama04}. However, closer to the Mott transition, the effects of changing effective Hubbard parameters with EEG conformation is likely to dominate over disorder effects. This was demonstrated experimentally in Ref. \cite{Hartmann14,Hartmann15}, and substantiated by ab-initio calculations of the band structure of fully EEG-ordered $\kappa$-salts in Ref. \cite{Guterding15}. In fact, it is possible to tune across the critical ratio $(U/W)_{\rm c}$ for the Mott transition in deuterated $\kappa$-(BEDT-TTF)$_2$Cu[N(CN)$_2$]Br by adjusting the cooling rate through $T_{\rm g}$. In such cooling-rate-dependent studies, important results, such as the critical slowing down of charge carriers at the Mott critical endpoint \cite{Hartmann15}, were made possible.

Whereas the control of disorder through the glassy freezing of the EEG disorder is reversible, it is hard to disentangle effects that result from disorder and from changes in effective Hubbard parameters, in particular when the material is situated close to the Mott MIT. Thus, the introduction of disorder by X-ray irradiation is a promising, complementary approach to study the influence of disorder, despite its irreversibility. A detailed review of the experimental procedures and various results are given in Ref.\,\cite{Sasaki12}.

Before turning to the results, it is important to summarize what type of disorder is created by X-ray irradiation. From studying molecular vibration modes in Cu-containing $\kappa$-phase charge-transfer salts by infrared optical spectroscopy \cite{Sasaki12b}, it was shown that irradiation has the strongest impact on vibrational models associated with the anion. This meets the intuitive expectation that X-rays should be mostly absorbed in the anion because they contain the heaviest atom, Cu. Thus, X-ray irradiation induces primarily molecular defects in the anion layer which creates a random lattice potential for the $\pi$ carriers in the BEDT-TTF layer. This notion was corroborated by DFT calculations \cite{Kang17}. In addition, there is no strong indication that X-ray irradiation changes the carrier concentration in the BEDT-TTF layer \cite{Sasaki12}. 

By now, there are various studies which have utilized X-ray irradiation to probe the interplay of disorder and correlations close to the Mott transition in $\kappa$-phase charge-transfer salts \cite{Analytis06,Sano10,Yoneyama10,Sasaki11,Sasaki12,Sasaki12b,Antal12,Sasaki15,Furukawa15,Gati18b,Urai19,Yamamoto20}. In fact, the corresponding Mott--Anderson model has attracted significant attention from a theoretical perspective, leading to a large number of numerical evaluations of its properties (e.g., Ref.\,\cite{Belitz94,Dobrosavljevic97,Shinaoka09,Chiesa08,Heidarian04,Pezzoli10,Tanaskovic03,Aguiar09,Aguiar13,Braganca15,Byczuk05,Lee16}). In the limit of independent electrons, the introduction of disorder is known to transform a metal into an Anderson insulator \cite{Anderson58,Lee85}. In the opposite limit of no disorder, but strong correlations, the latter will drive a metal--insulator transition. Naively, thus, both disorder and correlations promote insulating states. However, the interplay is much more complex, as discussed theoretically \cite{Radonjic10} and also witnessed experimentally in transport studies ($\rho(T)$) on $\kappa$-phase charge-transfer salts close to the Mott transition. Here, the impact of disorder on metallic salts on $\rho(T)$ \cite{Sano10,Sasaki11} is found to be distinct from those of the Mott insulators $\kappa$-Cl and $\kappa$-CuCN \cite{Sasaki12,Furukawa15}: For \mbox{$\kappa$-(BEDT-TTF)$_2$Cu[N(CN)$_2$]Br}, weak disorder increases the residual resistivity while simultaneously suppressing the superconducting critical temperature. For higher degrees of disorder, insulating behavior with d$\rho(T)/$d$T\,<\,0$ is observed across the full temperature range. On the contrary, for Mott insulators, whose $\rho(T)$ shows activation-type behavior before irradiation, $\rho(T)$ decreases at any given temperature with an increasing dose of irradiation. Generally, the impact of disorder is found to be stronger for $\kappa$-phase charge-transfer salts closer to the Mott transition.

These opposing trends of the effect of disorder on the metal vs. the Mott insulator motivate a detailed study of the Mott MIT in the presence of varying degrees of disorder and pressure. Such studies were performed in Refs.\,\cite{Gati18b,Urai19} and we show the resulting phase diagram of Ref.\,\cite{Gati18b} in Figure\,\ref{fig:shift-Mottline-disorder}b for irradiation times below 150\,h, corresponding to weak to moderate disorder. This study revealed a very clear tendency of the position and character of the MIT upon increasing disorder. While the MIT initially retains its discontinuous character at low temperatures, the location of the MIT is shifted to lower pressures and lower temperatures. Upon increasing irradiation further to $\sim$150\,h, no signatures of a discontinuous MIT were detected. Thus, for low disorder, the MIT shows characteristics of the pristine Mott transition. Only above a critical disorder strength is the Mott character no longer evident.

Interestingly, a lower critical pressure $p_{\rm cr}$ for increased levels of disorder strengths shows equivalent features to an increased critical correlation strength $(U/W)_{\rm c}$. This is consistent with theoretical predictions of a ``soft Coulomb gap'' \cite{Shinaoka09}---disorder widens the Hubbard bands by introducing a finite spectral weight at the Fermi level. Consequently, a larger $U$ is needed to fully open the Mott gap. This interpretation is supported by optical conductivity data on $\kappa$-(BEDT-TTF)$_2$Cu[N(CN)$_2$]Cl ($\kappa$-Cl) at ambient pressure \cite{Sasaki12} as well as from scanning tunneling microscopy studies of the normal-state density of states in $\kappa$-(BEDT-TTF)$_2$Cu[N(CN)$_2$]Br \cite{Diehl15},
even in its pristine form. 

Besides the position of the metal--insulator transition in the temperature--pressure phase diagram, the behavior of $\kappa$-Cl after longer irradiation at ambient and finite pressures also reveals interesting phenomena that motivate various further studies of disordered Mott systems. Possible questions of interest include how disorder affects the nature of the transformation of metallic into insulating regions across the first-order phase transition. The thermodynamic data of Ref.\,\cite{Gati18b,Gati17} clearly indicate an increased broadening of the jump-like changes in the volume across the first-order phase transition for samples with increasing levels of irradiation-induced disorder. NMR measurements on strongly irradiated $\kappa$-Cl under hydrostatic pressure \cite{Yamamoto20} revealed very slow dynamics, associated with an electronic Griffiths phase, located in proximity of the Mott phase boundary of the pristine, almost clean material. The authors of this work argued that the observed slow dynamics are incompatible with a macroscopically phase separated state and associated-domain wall motion. Further experimental studies will be useful in the future to support the intriguing physics presented above. 

In addition, it was suggested that the suppression of the first-order transition stabilizes the quantum-critical behavior of the Mott transition to lower temperatures \cite{Urai19}.  Furthermore, it was found that large irradiation times of $\sim$500\,h suppress the antiferromagnetic order~\cite{Furukawa15} that is present in pristine crystals at ambient pressure. The absence of long-range order in strongly disordered $\kappa$-Cl is particularly interesting in light of the discussion of the role of disorder in spin-liquid candidate systems. Clearly, an important step for the future is a detailed quantification of the X-ray-induced amount of disorder and its spatial distribution, which unfortunately is still lacking so far for the $\kappa$-phase charge-transfer salts. This input will be important for achieving quantitative comparisons between experiments and theories of disordered, correlated materials.

\subsection{Disorder in the Magnetic Phase: Scenario of a Valence Bond Solid Host with Orphan Spins}

\begin{figure}[H]
\begin{center}
\includegraphics[width=\columnwidth]{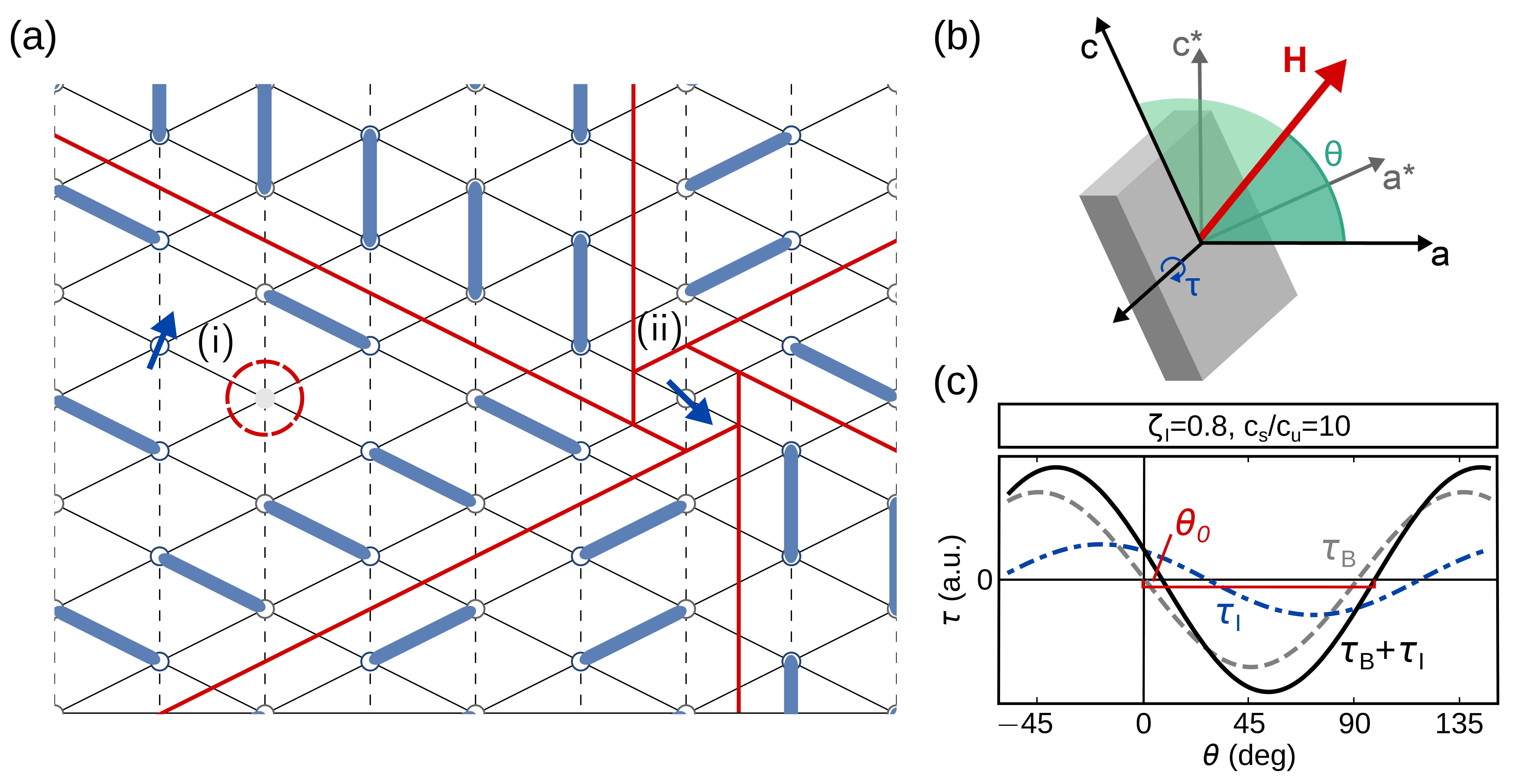} 
\end{center}
\caption{(\textbf{a}) Disorder scenarios in a valence-bond solid state, discussed for $\kappa$-(BEDT-TTF)$_2$Cu$_2$(CN)$_3$. Local spin 1/2, or so-called ``orphan spins'', may be caused by (i) vacancies in the anion layer, emphasized by the red circle or (ii) specific domain wall patterns, a possible result of randomness in the ethylene endgroup conformation. (\textbf{b}) Magnetic-torque setup, with crystal axes $a$ and $c$ (with $a^* \perp bc$, $c^* \perp ab$), magnetic field $\mathbf{H}$ and the angle $\theta$ between sample and field. (\textbf{c}) Theoretical magnetic torque $\tau$ dependence on field angle $\theta$, with indication of the angle shift $\theta_0$ and resolution of impurity ($\tau_{\rm I}$) and bulk ($\tau_{\rm B}$) contributions. Figures adapted from Ref.~\cite{Riedl19}.}
\label{fig:orphan-spins}
\end{figure}

Among the Mott insulating $\kappa$-phase charge-transfer salts, $\kappa$-(BEDT-TTF)$_2$Cu$_2$(CN)$_3$ \linebreak \mbox{($\kappa$-CuCN)} was recently at the center of increased interest in the context of disorder effects~\cite{Riedl19,Pustogow20,Miksch21,Matsuura22}. The magnetic ground state of this material is expected to have a vanishing net-magnetic moment, such as a quantum-spin-liquid (QSL) or valence-bond-solid (VBS) state. In these states, so-called
``orphan spins'' may emerge as a result of non-magnetic vacancies caused by defects in the anion layer, or by domain wall patterns in the VBS case, for instance, as a result of the randomized ethylene endgroup conformation discussed above (see Figure~\ref{fig:orphan-spins}a). 

The effects of orphan spins were argued in Ref.~\cite{Riedl19} to offer a consistent explanation for experimental observations in $\kappa$-CuCN such as magnetic torque~\cite{Isono16} and NMR~\cite{Shimizu06} measurements. This interpretation is in contrast to the originally proposed critical scenario~\cite{Pratt11,Winter17importance}, which was introduced based on the unconventional field-dependence of the $\mu$SR linewidth, i.e., of the magnetic susceptibility. In a critical scenario, one would expect either a uniform criticality or, as elaborated in Section~\ref{sec:non-Heisenberg}, in the presence of a DM interaction a staggered criticality. As detailed below, both scenarios are, however, not compatible with the magnetic-torque observations.

Theoretically, the magnetic torque can be expressed as $\tau = H^2 \chi_{\tau}(H) f(\theta)$, where $\chi_\tau(H)$ is the torque susceptibility with a possible non-trivial field dependence and $f(\theta)$ is a function of the angle $\theta$ between magnetic field and sample (see Figure~\ref{fig:orphan-spins}b). The experimental torque observations by Isono {et al.}~\cite{Isono16} can be summarized as:
(i) an unconventional field dependence of the torque susceptibility $\chi_\tau \propto H^{-0.8}$, (ii) a sinusoidal dependence of the torque as a function of magnetic-field angle $\tau \propto \sin(\theta - \theta_0)$, and (iii) a field-dependence of the angle shift $\theta_0(H)$. 
A critical scenario is not compatible with these three observations simultaneously. 
A uniform criticality would lead to sinusoidal-torque response for all relevant field strengths. This does not allow for a field-dependent angle-shift $\theta_0(H)$ as observed in the experiment. A staggered criticality, on the other hand, would lead to a sawtooth-shaped magnetic torque, also in contrast to the experimental observation. 

A consistent explanation for all three key torque features is given by the consideration of disorder-induced orphan spins. 
An orphan spin can be described by a localized magnetic moment and generally consists of a broad screening cloud, which depends on the interactions of the host system. The induced magnetic moment can then be described by a uniform contribution $\sum_{i^\prime\sim m}\langle \tilde{\mathbf{S}}_{i^\prime} \rangle = c_u \langle \tilde{\mathbf{S}}_{\text{I},m} \rangle$ and a staggered contribution $\sum_{i^\prime\sim m} \eta_{i^\prime}\langle \tilde{\mathbf{S}}_{i^\prime} \rangle 
= c_s \langle \tilde{\mathbf{S}}_{\text{I},m} \rangle$. Here, we labelled the sites surrounding the impurity site $m$ within the screening cloud with index $i^\prime$ and sublattice index $\eta_{i^\prime}=\pm 1$.
In contrast to the pristine bulk case, the induced staggered magnetic moment is then parallel to the induced defect magnetization. 
This ensures the impurity torque has sinusoidal field-dependence, even if the staggered contribution is the dominant one, allowing for a scenario which simultaneously leads to a sinusoidal magnetic-torque shape and a field-dependent angle shift. 
In Figure~\ref{fig:orphan-spins}c, a scenario of Ref.~\cite{Riedl19} is illustrated with the experimentally observed field exponent ($\zeta_{\rm I}=0.8$) and a dominant staggered contribution ($c_s/c_u=10$). The result is a sinusoidal angle dependence of the total torque $\tau=\tau_{\rm B}+\tau_{\rm I}$. The angle shift $\theta_0$ is field-dependent due to the relative shift in the bulk torque (with $\tau_{\rm B}\propto H^2$) and the impurity torque (with $\tau_{\rm I}\propto H^2 H^{-\zeta_{\rm I}}$). 

Considering expressions derived in the framework of finite-randomness large-spin fixed point (LSFP)~\cite{Westerberg95,Westerberg97}, it also turns out that the critical exponent is not as sample-dependent as one might expect for a disorder phenomenon. The corresponding exponent is restricted to be $2/3 < \zeta < 1$, so that the experimentally observed $\zeta=0.8$ falls in the middle of the possible range. 
These LSFP expressions also allow to predict the temperature and field dependence of the NMR linewidth $\nu$, in good agreement with experiment~\cite{Shimizu06}. 
They were recently employed in the analysis of a comparative study of NMR spin-relaxation-rate measurements for $\kappa$-phase compounds for which the magnetic ground state is argued not to order: \mbox{$\kappa$-CuCN}, $\kappa$-(BEDT-TTF)$_2$Ag$_2$(CN)$_3$ and \mbox{$\kappa$-(BEDT-TTF)$_2$Hg(SCN)$_2$Cl}~\cite{Pustogow20}.
Moreover, in $\kappa$-CuCN both, the measured NMR linewidth and magnetic torque become temperature independent around $\sim$1 K, suggesting a common impurity origin for NMR linewidth and magnetic torque. 
The VBS scenario was supported further recently by ESR measurements observing the opening of a spin gap below the $T^\ast$ anomaly~\cite{Miksch21}. The nature of this transition was investigated in inelastic neutron scattering (INS) measurements, where the observed crossover was interpreted in favor of a VBS instead of a QSL state~\cite{Matsuura22}.

In addition to offering a consistent explanation of the magnetic torque, NMR, ESR, and INS results, the orphan-spin scenario could potentially also solve an open issue for $\kappa$-CuCN which was subject for debate for over a decade. 
The linear temperature dependence of specific heat measurements~\cite{Yamashita08} was interpreted as the signature of a spinon Fermi surface, so that gapless spinon excitations of the QSL state would be present. Seemingly in contrast to this observation, thermal conductivity measurements~\cite{Yamashita09} revealed $\kappa/T=0$ for $T\rightarrow 0$, which is not compatible with the presence of low-lying fermionic excitations. Considering the presence of orphan spins, both of these observations are reasonable. In this scenario, low-lying excitations in the form of local domain walls may lead to a linear temperature dependence of specific heat. Since these excitations are local, they would not be observed in transport measurements such as thermal conductivity. 

\section{Conclusions and Outlook}
\label{sec:summary}

Organic charge-transfer salts, especially those of the $\kappa$-(BEDT-TTF)$_2X$ and $\kappa$-(BETS)$_2X$ family, are considered to be model systems to explore the physics of strongly correlated electron systems in proximity of the Mott metal--insulator transition. This notion has been corroborated by a successful description of a large set of properties of the systems in terms of the one-band Hubbard model in the strongly dimerized limit as well as the nice tunability of the real materials in laboratory settings. In recent years, it became clear that all salient features of these materials, however, can only be accurately captured when including further aspects in the generalized models. In this review, we covered a selection of new aspects that have proven to be relevant to the physics of $\kappa$-phase organic charge-transfer salts. In particular, we discussed theoretical and experimental evidence for the relevance of (i) magnetic interactions beyond the Heisenberg exchange, including spin-orbit and higher order ring-exchange couplings, (ii) intra-dimer degrees of freedom in creating novel states, such as electronically driven ferroelectric states and mixed superconducting order parameters, (iii) the coupling of correlated electrons to lattice degrees of freedom and (iv)~disorder. 
Nonetheless, a few open questions remain. Some examples are:

\begin{itemize}
    \item 
    As discussed in the review, ab-initio 
    extracted magnetic models for triangular $\kappa$-phase charge-transfer salts indicate the importance of four-spin interaction terms with spin-orbit coupling effects, as well as, in the presence of a magnetic field, possible 
    products of an odd number of spin operators, such as the scalar spin chirality.
    What (quantum) phases are to be expected arising from these interactions that are relevant for these~materials?
    \item How do intra-dimer charge and spin degrees of freedom conspire in the ground-state properties of frustrated Mott insulating $\kappa$-phase charge-transfer salts? In particular, what impact do the intra-dimer charge degrees of freedom have in promoting (impeding) the formation of long-range spin order? Can these effects be theoretically described with models containing both, spin and charge degrees of freedom?
    \item What is the role of magnetoelastic coupling in the formation of novel states of matter, such as putative spin-liquid states in $\kappa$-(BEDT-TTF)$_2X$? Can we develop accurate models to capture these effects?
    \item Can we quantitatively describe the impact of disorder on the properties of these charge-transfer salts close to the Mott metal--insulator transition experimentally and theoretically? To this end, how can we accurately quantify the level of disorder in real materials and determine the nature of the disorder and their spatial distribution? Which models and methods allow to theoretically describe the interaction between disorder and bulk properties properly?
    \item What novel phases may be realized under non-equilibrium conditions~\cite{buzzi2021}?
\end{itemize}

While this review focuses on the physics of the $\kappa$-phase organic charge-transfer salts specifically, many of the discussed aspects are relevant for the entire family of strongly correlated electron systems. For example, four-spin exchange interactions are expected to be relevant in 
triangular-lattice based inorganic Mott insulators with 4$d$ or 5$d$
transition metal ions. Likewise, as was pointed out in the past~\cite{Khomskii10}, charge effects might be generically relevant in frustrated Mott systems even deep in the Mott insulating phase. Recent examples are, for instance, the studies of the dielectric properties of the Kitaev magnet $\alpha$-RuCl$_3$~\cite{Mi22}. Furthermore, the coupling of the correlated electron system to the lattice is under intensive scrutiny in the study of, for instance, unconventional superconductors
(high-$T_c$ cuprates \cite{Vojta09}, Sr$_2$RuO$_4$ \cite{Mackenzie17} or Fe pnictides and chalcogenides~\cite{Fernandes14}),  frustrated magnets~\cite{Wosnitza16} and multiferroics \cite{vandenBrink08,spaldin2019}. Last but not least, the impact of disorder is now studied for instance in ultra-clean metals~\cite{Sunko20} and high-temperature superconductors~\cite{Cho18,Leroux19} by deliberately introducing defects through irradiation.

Many of the insights that we reviewed in this manuscript resulted from the continuous effort of advancing theoretical and experimental methods. With new techniques becoming available in the future, there is no doubt that the organic charge-transfer salts will remain key model systems for discovering, understanding and predicting physical properties arising from strong electron correlations in real materials.

\vspace{6pt} 

\authorcontributions{
All authors contributed to the conceptualization and writing of the manuscript.
All authors have read and agreed to the published version of the manuscript.
}

\funding{This work was funded by the Deutsche Forschungsgemeinschaft (DFG, German Research Foundation) for funding through TRR 288---422213477 (project A05). EG acknowledges support from the Max Planck Society.}

\dataavailability{Not applicable.
} 

\acknowledgments{We wish to thank all our collaborators and colleagues for many discussions on the physics of organic charge-transfer salts. 
}

\conflictsofinterest{The authors declare no conflict of interest.}

\begin{adjustwidth}{-\extralength}{0cm}
\reftitle{References}

\end{adjustwidth}
\end{document}